\documentclass[aps,prl,twocolumn,superscriptaddress,10pt]{revtex4-1}
\usepackage{graphicx}				% need for figures
\usepackage{epsfig}
\usepackage{amsmath,amsthm} % need for subequations
\usepackage{color}
\usepackage{verbatim}				% useful for program listings
\usepackage{ulem}
\begin{document}

\bibliographystyle{prlsty}

\title{Chirality in the Kagome Metal CsV$_3$Sb$_5$}

\author{H.J. Elmers}\email{elmers@uni-mainz.de}
\affiliation{Institut f\"{u}r Physik, Johannes Gutenberg-Universit\"{a}t, Staudingerweg 7, D-55128 Mainz, Germany}

\author{O. Tkach}
\affiliation{Institut f\"{u}r Physik, Johannes Gutenberg-Universit\"{a}t, Staudingerweg 7, D-55128 Mainz, Germany}
\affiliation{Sumy State University, Kharkivska 116, 40007 Sumy, Ukraine}

\author{Y. Lytvynenko}
\affiliation{Institut f\"{u}r Physik, Johannes Gutenberg-Universit\"{a}t, Staudingerweg 7, D-55128 Mainz, Germany}
\affiliation{Institute of Magnetism of the NAS and MES of Ukraine, 03142 Kyiv, Ukraine}

\author{P. Yogi}
\affiliation{Institut f\"{u}r Physik, Johannes Gutenberg-Universit\"{a}t, Staudingerweg 7, D-55128 Mainz, Germany}

\author{M. Schmitt}
\affiliation{Diamond Light Source Ltd., Didcot OX11 0DE, United Kingdom}
\affiliation{Physikalisches Institut and Würzburg-Dresden Cluster of Excellence ct.qmat, Julius-Maximilians-Universität, D-97074 Würzburg, Germany}

\author{D. Biswas}
\affiliation{Diamond Light Source Ltd., Didcot OX11 0DE, United Kingdom}

\author{J. Liu}
\affiliation{Diamond Light Source Ltd., Didcot OX11 0DE, United Kingdom}

\author{S. V. Chernov}
\affiliation{Deutsches Elektronen-Synchrotron DESY, 22607 Hamburg, Germany}

\author{M. Hoesch}
\affiliation{Deutsches Elektronen-Synchrotron DESY, 22607 Hamburg, Germany}

\author{D. Kutnyakhov}
\affiliation{Deutsches Elektronen-Synchrotron DESY, 22607 Hamburg, Germany}

\author{N. Wind}
\affiliation{Deutsches Elektronen-Synchrotron DESY, 22607 Hamburg, Germany}
\affiliation{Institut für Experimentelle und Angewandte Physik, Christian-Albrechts-Universität zu Kiel, 24098 Kiel, Germany}

\author{L. Wenthaus}
\affiliation{Deutsches Elektronen-Synchrotron DESY, 22607 Hamburg, Germany}

\author{M. Scholz}
\affiliation{Deutsches Elektronen-Synchrotron DESY, 22607 Hamburg, Germany}

\author{K.\,\,Rossnagel}
%\affiliation{Deutsches Elektronen-Synchrotron DESY, 22607 Hamburg, Germany}
\affiliation{Institut für Experimentelle und Angewandte Physik, Christian-Albrechts-Universität zu Kiel, 24098 Kiel, Germany}
\affiliation{Ruprecht Haensel Laboratory, Deutsches Elektronen-Synchrotron DESY, 22607 Hamburg, Germany}

\author{A. Gloskovskii}
\affiliation{Deutsches Elektronen-Synchrotron DESY, 22607 Hamburg, Germany}
\author{C. Schlueter}
\affiliation{Deutsches Elektronen-Synchrotron DESY, 22607 Hamburg, Germany}

\author{A. Winkelmann}
\affiliation{Academic Centre for Materials and Nanotechnology, 
AGH University of Krakow, 30059 Kraków, Poland
}

\author{A.-A. Haghighirad}
\affiliation{
Institute for Quantum Materials and Technologies, Karlsruhe Institute of Technology, 76021
Karlsruhe, Germany}

%\author{more authors?}
%\affiliation{more Instituts?}

\author{T.-L.\,\,Lee}
\affiliation{Diamond Light Source Ltd., Didcot OX11 0DE, United Kingdom}

\author{M. Sing}
\affiliation{Physikalisches Institut and Würzburg-Dresden Cluster of Excellence ct.qmat, Julius-Maximilians-Universität, D-97074 Würzburg, Germany}

\author{R. Claessen}
\affiliation{Physikalisches Institut and Würzburg-Dresden Cluster of Excellence ct.qmat, Julius-Maximilians-Universität, D-97074 Würzburg, Germany}

\author{M. Le Tacon}
\affiliation{
Institute for Quantum Materials and Technologies, Karlsruhe Institute of Technology, 76021
Karlsruhe, Germany}

\author{J. Demsar}
\affiliation{Institut f\"{u}r Physik, Johannes Gutenberg-Universit\"{a}t, Staudingerweg 7, D-55128 Mainz, Germany}

\author{G. Sch{\"o}nhense}
\affiliation{Institut f\"{u}r Physik, Johannes Gutenberg-Universit\"{a}t, Staudingerweg 7, D-55128 Mainz, Germany}

\author{O. Fedchenko}
\affiliation{Institut f\"{u}r Physik, Johannes Gutenberg-Universit\"{a}t, Staudingerweg 7, D-55128 Mainz, Germany}

%\pacs{79.60-i, 73.20-r, 75.25-j, 75.70-i}

\date{\today}

\begin{abstract}

Using x-ray photoelectron diffraction (XPD) and angle-resolved photoemission spectroscopy, we study 
photoemission intensity changes related to changes in the geometric and electronic structure   
%structural and electronic band structure changes 
in the kagome metal CsV$_3$Sb$_5$ upon transition to an unconventional charge density wave (CDW) state. The XPD patterns reveal the presence of a chiral atomic structure in the CDW phase. Furthermore, using circularly polarized x-rays, we have found a pronounced non-trivial circular dichroism in the angular distribution of the valence band photoemission in the CDW phase, indicating a chirality of the electronic structure. This observation is consistent with the proposed orbital loop current order. In view of a negligible spontaneous Kerr signal in recent magneto-optical studies, the results suggest an antiferromagnetic coupling of the orbital magnetic moments along the $c$-axis. While the inherent structural chirality may also induce circular dichroism, the observed asymmetry values seem to be too large in the case of the weak structural distortions caused by the CDW.

\end{abstract}

\maketitle

%\section{Introduction}

Kagome systems have recently attracted interest because of their  unique electronic band structure, which exhibits delocalized
electrons,  Dirac points, flat
bands, and multiple van Hove singularities (vHS) near
the Fermi level~\cite{Guo2024,Hu2023,Kang2022,Sato2017}. 
The large density of states near vHSs can promote various electronic orders, such as superconductivity, chiral charge density wave (CDW), and orbital loop current order~\cite{Xing2024,Xu2015,Guo2009,Tang2011}.
The vHSs in combination with flat bands can be the origin of correlated many-body ground states~\cite{Wang2013,Kiesel2013,Kiesel2012,
Yu2012,Mielke1992,Vonsovsky1989}. 
In particular, 
the kagome metals AV$_3$Sb$_5$ (A = K, Rb, Cs) 
combine unconventional charge orders~\cite{Frachet2024,Xu2022,Nie2022,Zhao2021,Li2021,Liang2021,Chen2021}, superconductivity~\cite{Guguchia2023,Jiang2022,Zhang2021,Neupert2021,Chen2021a,Chen2021b,Tan2021,Ortiz2020}, lattice frustration, and non-trivial topology~\cite{Mielke2022,Jiang2021,Yang2020}.
Spectroscopic evidence for topological properties and correlation effects have been provided by ARPES studies~\cite{Ortiz2020,Hu2022} and density functional theory~\cite{Lin2021,Park2021,Feng2021,Wu2021,Denner2021}. In particular, the charge order has been related to time-reversal symmetry breaking (TRS)~\cite{Guguchia2023,Xu2022,Mielke2022,Jiang2021} or
rotational symmetry breaking~\cite{Nie2022,Chen2021,Zhao2021}.

In the normal high temperature state, AV$_3$Sb$_5$ is a topological
metal~\cite{Ortiz2021,Ortiz2020} with topologically protected surface
states near the Fermi level~\cite{Ortiz2020}. In the low temperature CDW state, scanning tunneling spectroscopy (STM)~\cite{Jiang2021} and theory~\cite{Feng2021} have suggested a chiral CDW order that breaks the time-reversal symmetry~\cite{Feng2021}. 
%However, the link between the
%chiral CDW order and the symmetry breaking is still an open question.
Observations of an
anomalous Hall effect~\cite{Wei2024,Zhou2022,Yu2021,Yang2020} together with
muon spin relaxation~\cite{Mielke2022,Khasanov2022} and
magneto-optical Kerr effect (MOKE) studies~\cite{Hu2022a,Wu2022,Xu2022} provided further evidence for the spontaneous time-reversal symmetry breaking. 
However, recent dedicated MOKE studies with zero-loop Sagnac interferometers~\cite{Farhang2023,Wang2024,Saykin2023} and STM~\cite{Li2022} conclude that the occurrence of time-reversal symmetry breaking is unlikely.

Here, we use electron momentum microscopy to gain further insight into the structural and electronic chiral order in this system. Hard x-ray photoelectron diffraction (XPD) reveals a chiral order in the atomic structure in the CDW phase. On the other hand, photoelectrons excited by circularly polarized light in the soft x-ray region show spontaneous chiral symmetry breaking in the valence band states, which is particularly pronounced at certain mirror symmetry points in reciprocal space ({\it i.e.} M-points).
We discuss time-reversal symmetry breaking or mirror symmetry breaking as possible origins for the observed effects.

%\section{Experimental}

The experiments were performed on single crystals of CsV$_3$Sb$_5$ grown by the flux method and characterized by x-ray diffraction and energy-dispersive x-ray analysis~\cite{Frachet2024,Stier2024}.
The CsV$_3$Sb$_5$ structure and the experimental geometries for the photoemission experiments are given in the Supplemental Material (SM)~\cite{Suppl}.
%shown in Figs.~\ref{Fig1}(a) and \ref{Fig1}(b), respectively. CsV$_3$Sb$_5$ consists of a planar arrangement of V atoms forming a kagome lattice consisting of three sets of parallel lines of V atoms (red) with Sb atoms (gray) as nearest neighbors. The V-Sb planes are separated by Cs atoms, thus forming an electronically two-dimensional lattice. 
The single crystals were freshly cleaved in ultrahigh vacuum.

%\section{Experimental Results}
%\begin{figure}
%\includegraphics[width=\columnwidth]{FigureSchematics1_v2.pdf}
%\caption{\label{Fig1} 
%(a) Unit cell of CsV$_3$Sb$_5$ in the normal state.
%(b) Experimental geometry. The circularly polarized x-rays impigne on the (001) %surface at an angle $\theta$ within the $b-c$ plane or 90$^{\circ}$ rotated.
%(c) Brillouin zone of CsV$_3$Sb$_5$ in the high temperature phase with marked high-%symmetry points. The high-temperature phase notation was used throughout the %paper).
%(sample azimuth rotated by 90 degrees).
%(c) Single plane of the V atoms constituting the Kagome lattice. Small arrows indicate the loop current order. The resulting orbital moments (blue arrows) form a 2x2 superstructure. The corresponding Wigner-Seitz cell is indicated in green as compared to the normal state cell indicated in red. Red and green arrows indicate the corresponding unit vectors along the $a$-axis.
%(d) Reciprocal space representation of the Fermi surface.
%(grey level is proportional to the photoemission intensity). 
%Green and red hexagons indicate the Brillouin zones of the %normal state structure (red) and the 2x2 superstructure (green).
%Green and red arrows mark the corresponding reciprocal lattice vectors. The blue arrows indicate the expected orbital moments near the M-points where the Fermi wave vector and the $\Gamma$ points of the 2x2 superstructure coincide.
%}
%\end{figure}

CsV$_3$Sb$_5$ shows a CDW reconstruction below the transition temperature ($T_{\rm CDW}=94$~K)~\cite{Hu2023}. To study the CDW-induced changes in the crystal structure we used core level XPD. Experiments in the hard x-ray range at 6~keV were performed at the time-of-flight momentum microscopy endstation of the hard x-ray beamline P22 at PETRA~III~\cite{Schlueter2019}. The angle of incidence was $\theta=10^{\circ}$ with respect to the sample plane and the energy resolution was set to 600~meV.
XPD was performed on the Cs $4d$, V $2p$, and Sb $3d$ core levels, all of which show a pronounced XPD pattern with a sixfold symmetry.

\begin{figure}
\includegraphics[width=\columnwidth]{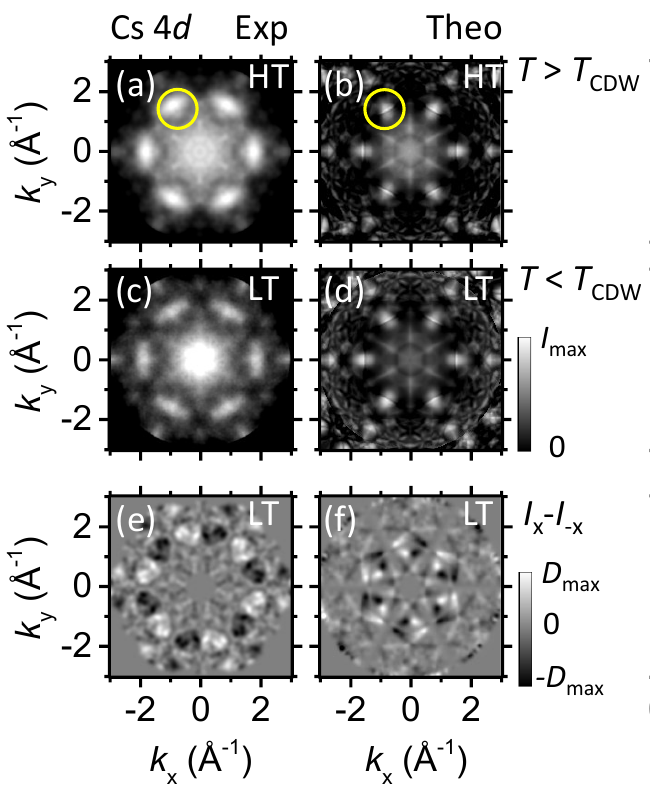}
\caption{\label{Fig5} 
X-ray photoelectron diffraction of CsV$_3$Sb$_5$ measured at a photon energy of 6~keV for $T>T_{\rm CDW}$ (HT, 115~K) and $T<T_{\rm CDW}$ (LT, 30~K). 
Comparison of  experimental (a,c) and theoretical  (b,d) data. 
%Yellow circles indicate good agreement between theoretical and experimental data. 
(e) Difference between the original data from (c) and the same data but mirrored at $k_x=0$ to highlight the broken mirror symmetry. (f) Similar data for the calculated results (see SM~\cite{Suppl} for HT and other core levels).
}
\end{figure}

The Cs $4d$ pattern [Fig.~\ref{Fig5}(a)] recorded in the high-T phase shows an inner sixfold star (vertical orientation) surrounded by a ring of high intensity peaks in a straddled orientation (yellow circle). This is consistent with the pattern [Fig.~\ref{Fig5}(b)]  calculated  using a Bloch wave approach with the normal state crystal structure parameters determined by XRD ~\cite{Kautzsch2023,Ortiz2021a}.
Below the CDW transition, the overall appearance of the XPD pattern does not change much 
%but the inner star has a lower intensity 
[Fig.~\ref{Fig5}(c)], in agreement with the Bloch wave calculation considering the $(2\times 2\times 4)$ reconstruction~\cite{Kautzsch2023,Ortiz2021a} [Fig.~\ref{Fig5}(d)].
To highlight the broken mirror symmetry (chirality) of the low-T XPD pattern we plot the difference $I(k_x,k_y)-I(-k_x,k_y)$ in Fig.~\ref{Fig5}(e). 
While the difference still shows a six-fold symmetry, the mirror symmetry of the XPD pattern is broken, as shown by a counterclockwise bending of the tips of the inner star in the XPD pattern. Indeed, a similar chirality shows up in the calculated XPD pattern [Fig.~\ref{Fig5}(f)] based on the low-T XRD data from Refs.~\cite{Kautzsch2023,Ortiz2021a}. A corresponding analysis was performed for the V $2p$ and Sb $3d$ core level XPD patterns for $T<T_{\rm CDW}$ and $T>T_{\rm CDW}$ (see SM~\cite{Suppl}).
%The SM includes additional references [xx-xy].
These results demonstrate the structural chirality of the CDW order and imply that the domain size is larger than the x-ray footprint on the sample ($\approx$ 50 $\mu$m in diameter), consistent with earlier domain mapping studies ~\cite{Xu2022}.
%This indicates that the mirror symmetry of the normal state is broken in the CDW phase.
%.a breaking of the structural mirror symmetry in the normal state that is due to the 3-dimensional CDW phase.

\begin{figure*}
\includegraphics[width=\textwidth]{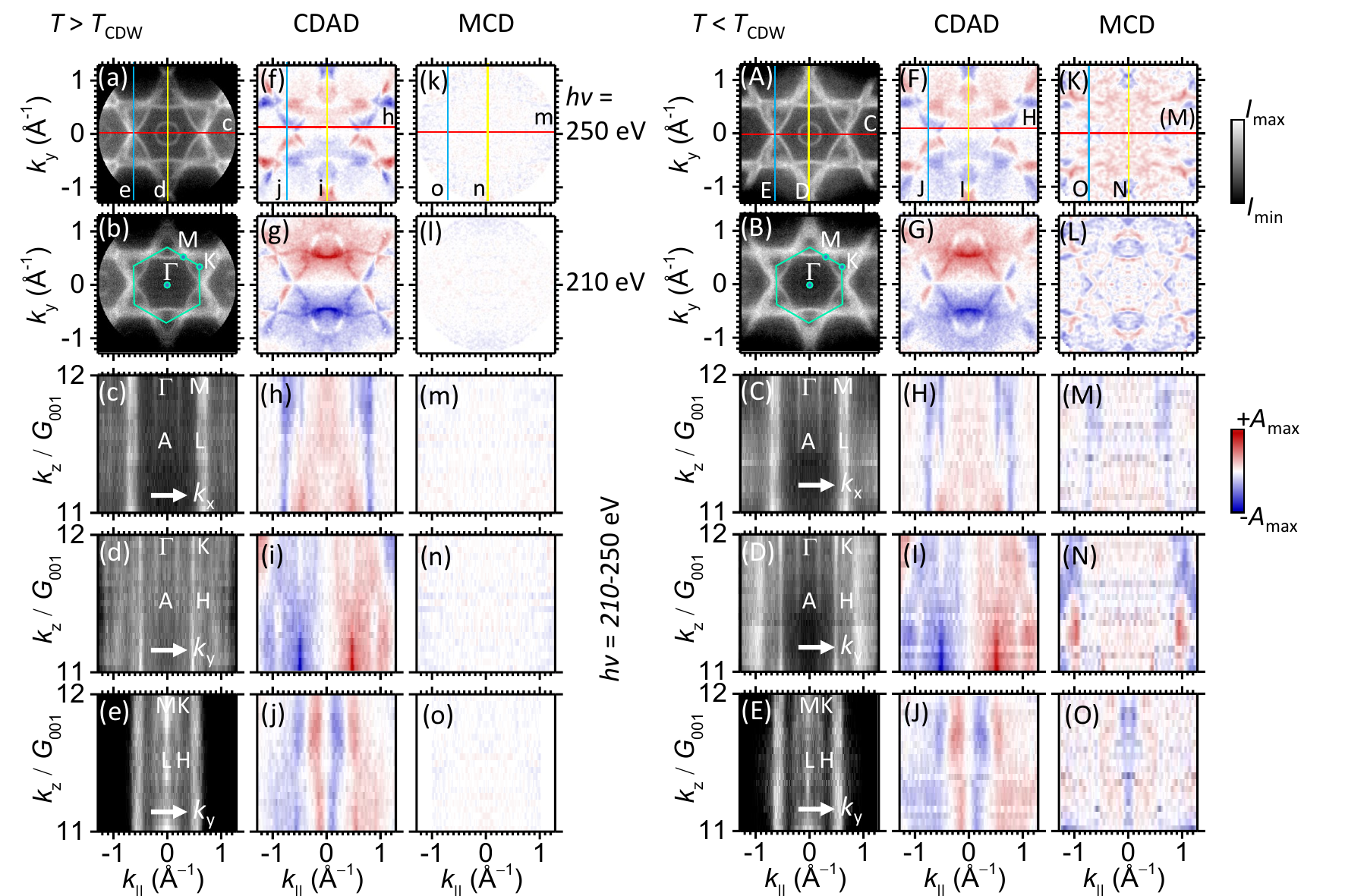}
\caption{\label{Fig2} 
(a,b) Fermi surface of CsV$_3$Sb$_5$ measured at $h\nu=250$~eV (a) and 210~eV (b) for $T>T_{\rm CDW}$. The photon incidence is from the right.
(c-e) Fermi surface sections in the $k_z - k_{||}$ planes (perpendicular to the surface) corresponding to the lines indicated in (a).
The photon energy has been varied between 210 and 250~eV in small steps.
(f-j) $A_{\rm CDAD}$ in the same Fermi surface sections as in (a-e) [(h) shifted to $k_y>0$ to avoid the $A_{\rm CDAD}=0$-line]. The maximum asymmetry of the color scale (right) is $A_{\rm max}=0.5$.
(k-o) $A_{\rm MCD}$ in the same Fermi surface sections as in (a-e). The maximum asymmetry of the color scale is in this case $A_{\rm max}=0.1$.
(A-O) Similar data measured at 30~K ($T<T_{\rm CDW}$) plotted on the same scales.
}
\end{figure*}

%\begin{figure*}
%\includegraphics[width=\textwidth]{FigureDiamond_bands_v1.pdf}
%\caption{\label{Fig3} 
%(a-d) Temperature dependence of band dispersion. Constant energy sections of the %photoemission intensities in the $k_x - k_y$ plane at the indicated binding energies, %measured at a photon energy of 250~eV. 
%(e-h) Band dispersions along the indicated high symmetry directions in reciprocal space.
%Data on the left are measured for $T>T_{CDW}$ and on the right for $T<T_{CDW}$. 
%}
%\end{figure*}

To detect the chirality in the electronic structure valence band photoemission was performed. 
Circular dichroism experiments in the soft x-ray range were performed 
at the soft x-ray ARPES endstation of Beamline I09 at Diamond Light Source, 
UK~\cite{Schmitt2024}.
%at the endstation of the soft x-ray branch of beamline I09 at the Diamond Light Source,  
%The electron analyzer was a large single hemispherical spectrometer. 
The angle of incidence for the circularly polarized x-rays was $\theta = 22.5^{\circ}$ with respect to the sample surface, 
which was oriented to align the $\Gamma$-M-L plane with the incident beam. 
The total energy resolution was set to 50~meV. 
Figure~\ref{Fig2} shows the experimental Fermi surface of CsV$_3$Sb$_5$ measured at the two temperatures $T_H=115$~K and $T_L=30$~K, {\it i.e.}, above and below $T_{\rm CDW}=94$~K.
In Figs.~\ref{Fig2}(a-e) we plot the photoemission intensity distribution $I(E_F,k_x,k_y)$ at the Fermi energy as a function of  the parallel momentum $k_x$, $k_y$. The photon energies were varied in an interval from 210~eV to 250~eV, covering a perpendicular momentum range from $k_z=11 G_{001}$ to $12 G_{001}$
 [Fig.~\ref{Fig2}(c-e)]. We observe no significant dispersion along $k_z$, indicating a two-dimensional character of the electronic states. This is in agreement with previous observations~\cite{Hu2023}.

From the photoemission intensities measured with circularly polarized x-ray excitation, $I^{+/-}$, we determine the asymmetry  
%of the circular dichroism in the angular distribution (CDAD), 
$A^*=(I^+-I^-)/(I^++I^-)$. 
The non-relativistic circular dichroism in the angular distribution (CDAD) is  a geometric effect and it is strictly antisymmetric with respect to the photon plane of incidence, which coincides with the $\Gamma$-M-A crystal mirror plane~\cite{Westphal1989}.
To separate the CDAD from any other contribution, we calculate the asymmetry shown in Figs.~\ref{Fig2}(f-j) as 
$A_{\rm CDAD}(k_x,k_y)=(1/2)[A^*(k_x,k_y)-A^*(k_x,-k_y)]$
(see detailed discussion of the separation procedure in Ref.~\cite{Fedchenko2024}).
The maximum asymmetry is 0.5, and leads to a significant intensity redistribution when the x-ray helicity is switched.
The CDAD asymmetry depends strongly on the photon energy, as can be seen by comparing Figs.~\ref{Fig2}(f) and \ref{Fig2}(g) (see Ref.~\cite{Fedchenko2019}). The photon energy series shows a continuous change of the asymmetry with varying $k_z$ for several electronic bands. Only the states near the M-L-line show a constant asymmetry independent of the photon energy, as emphasized by the $k_z$-dependence shown in Figs.~\ref{Fig2}(h) and \ref{Fig2}(j).

Note that at $T>T_{\rm CDW}$, the dichroism is completely dominated by the CDAD. This can be visualized by the complementary magnetic circular dichroism (MCD), which is symmetric with respect to the $k_y$ direction~\cite{Bansmann1992},
$A_{\rm MCD}(k_x,k_y)=(1/2)[A^*(k_x,k_y)+A^*(k_x,-k_y)]$,
as shown in Figs.~\ref{Fig2}(k-o). $A_{\rm MCD}$ is less than 0.01 and shows no systematic variation as a function of parallel and perpendicular 
momentum. The lack of $A_{\rm MCD}$ confirms the non-magnetic character of the electronic states in CsV$_3$Sb$_5$ in its normal state at $T>T_{\rm CDW}$. 

At 30~K, the average photoemission intensity distribution shown in Figs.~\ref{Fig2}(A-E) has not changed much. There are no obvious changes in the dispersion of the visible bands, only the intensity distribution near the M-L line [central line at $k_{||}=0$ in Fig.~\ref{Fig2}(E)] changes slightly.
This observation is in agreement with previous reports using a higher energy resolution~\cite{Nakayama2021,Azoury2023}, where the CDW-induced gap opening appears near the M-point. The CDW-induced band gap is discussed in the SM \cite{Suppl}.

The CDAD asymmetry in the CDW state [Figs.~\ref{Fig2}(F-J)] shows a similar pattern to that observed in the normal state. The fact that no obvious CDW-induced change in CDAD is observed implies the absence of a significant change in orbital charge redistribution. 

In contrast, the low-T data show a significant $k_y$-symmetric $A_{\rm MCD}$
 [Figs.~\ref{Fig2}(K-O)]. A prominent, almost $k_z$-independent, contribution appears near the M-L line at $(k_x=\pm 0.7$~\AA$^{-1}$, $k_y=0, k_z)$. This is unexpected for a paramagnetic system. The two M-points at $k_y=0$ coincide with the antisymmetry axis of the CDAD, where the CDAD vanishes. The large asymmetry of about -0.1 is essentially independent of the photon energy [Figs.~\ref{Fig2}(M) and \ref{Fig2}(O)]. 
Especially near $k_y=0$ a pronounced non-zero $A_{\rm MCD}$ occurs that cannot be explained by an artefact due to different sample illumination.
Significant $A_{\rm MCD}$ values also show up at $(k_x=0$, $k_y=\pm 1$~\AA$^{-1},k_z)$
%$(k_x,k_y)=(0,\pm 1 {\rm \AA}^{-1})$
[Fig.~\ref{Fig2}(N)]. Here, the sign of  
$A_{\rm MCD}$ changes at about $k_z=11.5G_{001}$, which hints to additional diffraction-related asymmetries or matrix-element effects~\cite{Fedchenko2019}.
%could be an indication of a residual CDAD caused by an altered vignitation due to a movement of the photon footprint between left and right polarization.

Further details of the band dispersion versus binding energy as measured at 30~K and 115~K are discussed in SM~\cite{Suppl}.

\begin{figure}
\includegraphics[width=\columnwidth]{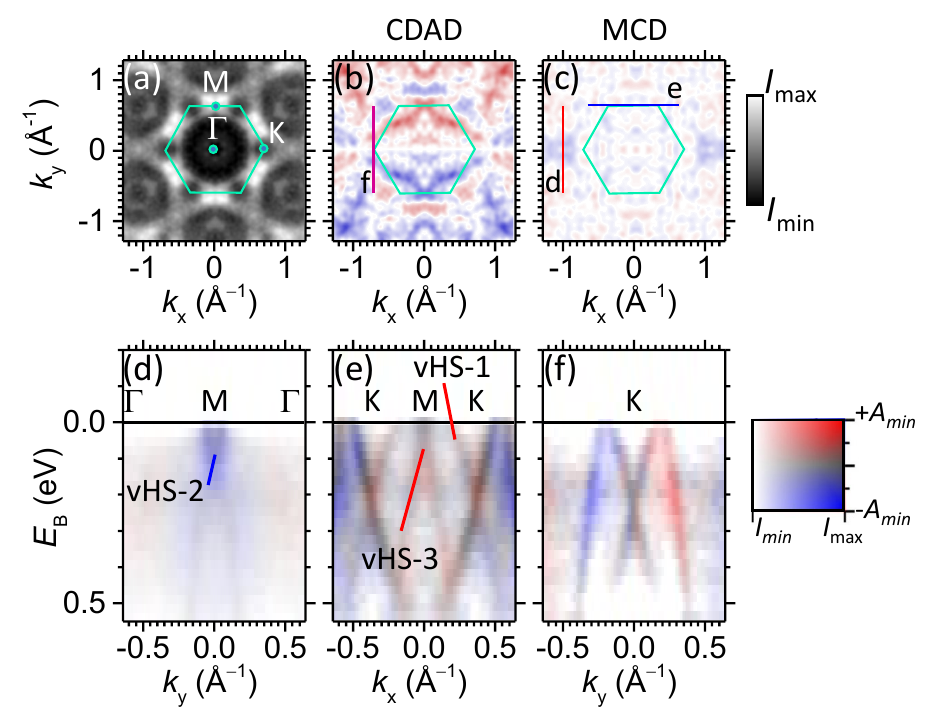}
\caption{\label{Fig4} 
(a) Fermi surface of CsV$_3$Sb$_5$ measured at a photon energy of 330~eV for $T<T_{\rm CDW}$. The photon incidence is from the right.
(b) CDAD texture of the Fermi surface. The maximum asymmetry of the color scale (right) is $A_{\rm max}=0.5$. 
(c) MCD asymmetry of the Fermi surface.
The maximum asymmetry of the color scale (right) is $A_{\rm max}=0.1$.
(d,e) Band dispersion with overlaid MCD asymmetry (two-dimensional color scale on the right) along the indicated lines in (c).
(f) Band dispersion with overlaid CDAD asymmetry across the K-point along the indicated line in (b). 
}
\end{figure}

To further elucidate the CDW-induced chirality using a different experimental geometry, we performed a soft x-ray photoemission experiment with the time-of-flight momentum microscope at the soft x-ray beamline P04 at PETRA III, DESY, Germany~\cite{Tkach2024}, with the total energy resolution set to 34~meV. We used 330~eV x-rays (corresponding to $k_z=13.7G_{001}$) and the same angle of incidence $\theta = 22.5^{\circ}$. In this case, however, the plane of incidence was along the $\Gamma$-K-H plane. 

\begin{figure}
\includegraphics[width=\columnwidth]{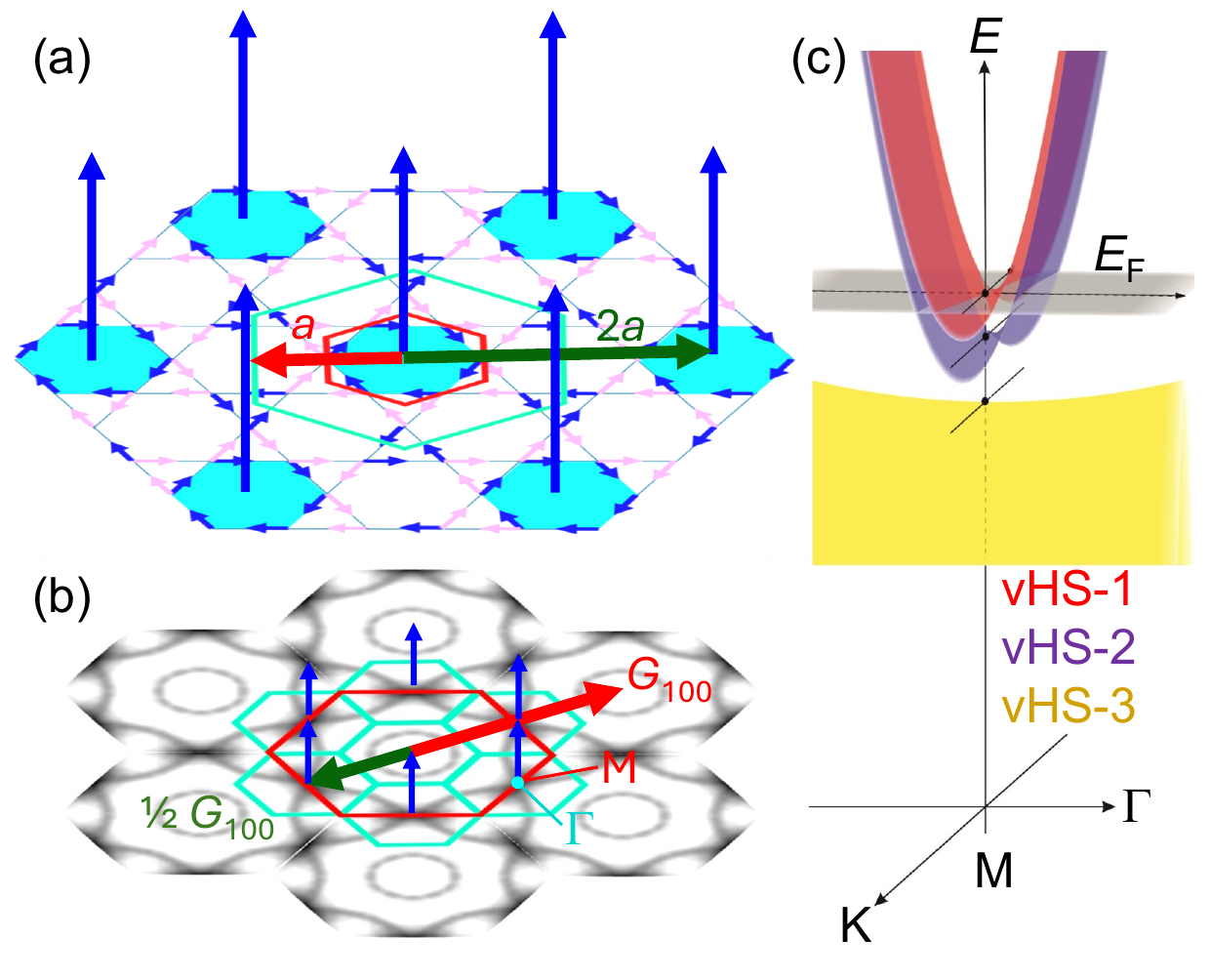}
\caption{\label{Figdiscuss} 
(a) V-Sb plane constituting the kagome lattice. Small arrows indicate the loop current order forming a $2\times2$  superstructure
of orbital moments (large arrows). 
The superstructure Wigner-Seitz cell is indicated in green and the normal state unit cell in red with corresponding unit vectors along the $a$-axis.
(b) Fermi surface in reciprocal space
(grey level is proportional to the measured photoemission intensity). 
Hexagons indicate the Brillouin zones of the 
normal state structure (red) and the $2\times2$  superstructure (green) with
corresponding reciprocal lattice vectors. 
The blue arrows indicate the expected orbital moments near the M-points where the Fermi wave vector and the $\Gamma$ points of the $2\times2$  superstructure coincide.
(c) Sketch of the first three vHSs near $E_F$.
}
\end{figure}

Figure~\ref{Fig4}(a) shows the measured Fermi surface recorded in this configuration at 30~K. 
The CDAD asymmetry at the Fermi surface, Fig.~\ref{Fig4}(b) is antisymmetric with respect to the plane of incidence, with the maximum asymmetry $A_{\rm max}$ $\approx$ 0.5 similar to the previous case. The MCD asymmetry Fig.~\ref{Fig4}(c) also shows negative values up to $-0.1$ near the M-point at $(k_x,k_y)=(\pm 1$~\AA$^{-1},0)$ in the adjacent Brillouin zones on the $\Gamma$-K symmetry axis ($k_y=0$-axis).
The sections along M-$\Gamma$ [Fig.~\ref{Fig4}(d)] and 
M-K [Fig.~\ref{Fig4}(e)] suggest that the negative MCD asymmetry
is related to the electron-like band with a maximum binding energy of 0.1~eV at the M-point, which belongs to the van Hove singularity vHS-2 [see SM for a detailed description of the vHSs~\cite{Suppl} and Fig.~\ref{Figdiscuss}(c)]. 
%Negative values for $A_{\rm MCD}$ also appear in the region of the
%$\gamma$-band belonging to vHS-4~\cite{Hu2022} at binding energies of 0.2~eV [left and right edges in Fig.~\ref{Fig4}(e)]. 
Bands associated with vHS-1, {\it i.e.}, the flat band near the Fermi level at the M-point [see Fig.~\ref{Fig4}(e)], and vHS-3 (hole-like band)
have positive $A_{\rm MCD}$ values. The large MCD asymmetries near the M-points confirm our findings obtained with the photon plane of incidence coinciding with the $\Gamma$-M-A plane [Fig.~\ref{Fig2}]. 

%As a side note, the existence of a Dirac cone at the K-point is confirmed by the
%CDAD asymmetry shown in Fig.~\ref{Fig4}(f): The two bands crossing the Dirac point %at a binding energy of 275~meV show a positive (negative) asymmetry at positive %(negative) $k_y$ above (below) the Dirac point. This is the expected behavior for a %Dirac point.

%\section{Discussion}

%\section{Conclusion}

%Two alternative models for kagome systems could explain the observed electronic chirality.
Since the multiple vHSs with opposite mirror eigenvalues are close in energy, it has been proposed that the nearest neighbor electron repulsion favors a ground state with coexisting loop current order and charge-bond order~\cite{Li2024}. The loop current order imposes a time-reversal symmetry breaking in the CDW phase.

%\section{Experimental Results}

Figure~\ref{Figdiscuss}(a) schematically shows the charge loop current as adapted from Ref.~\cite{Hu2023}. It can be visualized by alternating charge currents along the three sets of parallel lines connecting the V atom positions.  Thus, the charge current vanishes when integrated over the unit cell. The ordered charge current along the V-V bonds leads to loop currents, [Fig.~\ref{Figdiscuss}(a)], that bypass the six V atoms. The current direction of individual bonds can be either clockwise (blue arrows) or counterclockwise (pink arrows). Every second loop shows an unfrustrated loop current (shown in light blue) forming a $2\times2$ superlattice. %The charge loop current results in an orbital magnetic moment that breaks the time-reversal symmetry.
%The time-reversal symmetry breaking is attributed to the strong electron-electron interaction of the hybridized frustrated state formed by the V 3$d$ orbitals, which form van Hove singularities near the M-points.

In reciprocal space, the corresponding Brillouin zone of the $2\times2$ structure is shown in green in Fig.~\ref{Figdiscuss}(b). Here, some $\Gamma$-points of the superstructure Brillouin zone coincide with the M-points of the larger normal-state Brillouin zone [red in Fig.~\ref{Figdiscuss}(b)].
Since the van Hove singularities near the Fermi surface are located at these M-points, states with finite orbital angular momentum causing a MCD are also expected at these coincidence points [blue arrows in Fig.~\ref{Figdiscuss}(b)].
%This prediction is in good agreement with out experimental results.

Our experimental results are thus consistent with recent reports on the anomalous Hall effect ~\cite{Wei2024,Zhou2022,Yu2021,Yang2020}, observations of chiral charge order in STM studies \cite{Jiang2021} and the early magneto-optical Kerr effect studies ~\cite{Wei2024,Zhou2022,Yu2021,Yang2020,Mielke2022,Khasanov2022,Hu2022a,Wu2022,Xu2022}, all suggesting an orbital loop-current-order. On the other hand, recent dedicated polar Kerr effect measurements seem to rule out the time-reversal symmetry breaking in CsV$_3$Sb$_5$~\cite{Saykin2023,Farhang2023}. The absence of a measurable spontaneous Kerr effect (bulk probe) ~\cite{Saykin2023}, with strong evidence for chiral charge ordering from surface sensitive techniques such as STM \cite{Jiang2021} and the presented ARPES data, may imply an antiferromagnetic interlayer ordering of the orbital magnetic moments in CsV$_3$Sb$_5$. 
In this case, optical dichroism averaged over many atomic layers would disappear.

On the other hand, our photoelectron diffraction results at 6~keV have shown
have shown that the lattice exhibits a chiral structure in the CDW state,
which is consistent with XRD data implying a stacking of different CDW patterns along the
$c$-axis~\cite{Kautzsch2023,Ortiz2021a}. The coexistence of the Star-of-David and inverse Star-of-David reconstructions in the CDW state is demonstrated by XRD~\cite{Xiao2023,Ortiz2021}, nuclear quadrupole resonance (NQR)~\cite{Feng2023}, and nuclear magnetic resonance (NMR)~\cite{Frassineti2023} measurements.
Such structural chirality may also induce the observed valence-band MCD effect, which is in this case better described as an intrinsic instead of a magnetic circular dichroism~\cite{Brinkman2024}. 
%Distinguishing between the two models requires an approach similar to that of Ref.~\cite{Saykin2023}, which is not easy for photoemission experiments. 
In fact, a three-dimensional structural and electronic chirality of the helix-type without time-reversal symmetry breaking would also be consistent with our experimental results. 
The observed asymmetry values of the order of 0.1 are, however, much larger than previously assumed for the natural circular dichroism.
%, especially considering the rather small CDW induced changes in atomic positions. 
%However, previously observed natural circular dichoism effects were much smaller than the ones observed here.

%Photoemission calculations based on ab-initio theory could shed light on the origin of the chirality.
%A detailed temperature dependence of XPD and MCD will be needed to clarify whether $T_{\rm CDW}$ coincides with the onset of MCD.
%Spectroscopic experiments under $c$-axis compression are also a promising way to elucidate the microscopic origin of the observed chirality in AV$_3$Sb$_5$.

\hspace{1cm}

\begin{acknowledgments}
This work was funded by the Deutsche Forschungsgemeinschaft (DFG, German Research Foundation), grant no. TRR288–422213477 (projects B03, B04, and B08),
and grant no Scho341/16-1, 
through the Collaborative Research Center SFB 1170 ToCoTronics (Project No. 258499086), through the Würzburg–Dresden Cluster of Excellence on Complexity and Topology in Quantum Matter ct.qmat (EXC 2147, Project No. 390858490),
and by the BMBF (projects 05K22UM2, 05K22UM4, and 05K22WW1). 
Funding for the instrument by the Federal Ministry of Education and Research (BMBF) under framework program ErUM is gratefully acknowledged. We thank the Diamond Light Source for providing beamtime at beamline I09.
We acknowledge DESY (Hamburg, Germany) for the provision of experimental facilities. Parts of this research were carried out at PETRA III using beamlines P04 and P22.  
\end{acknowledgments}

\bibliographystyle{apsrev4-1}

%\bibliographystyle{apsrev4-2}
%\bibliography{CsV3Sb5}

\begin{thebibliography}{67}%
\makeatletter
\providecommand \@ifxundefined [1]{%
 \@ifx{#1\undefined}
}%
\providecommand \@ifnum [1]{%
 \ifnum #1\expandafter \@firstoftwo
 \else \expandafter \@secondoftwo
 \fi
}%
\providecommand \@ifx [1]{%
 \ifx #1\expandafter \@firstoftwo
 \else \expandafter \@secondoftwo
 \fi
}%
\providecommand \natexlab [1]{#1}%
\providecommand \enquote  [1]{``#1''}%
\providecommand \bibnamefont  [1]{#1}%
\providecommand \bibfnamefont [1]{#1}%
\providecommand \citenamefont [1]{#1}%
\providecommand \href@noop [0]{\@secondoftwo}%
\providecommand \href [0]{\begingroup \@sanitize@url \@href}%
\providecommand \@href[1]{\@@startlink{#1}\@@href}%
\providecommand \@@href[1]{\endgroup#1\@@endlink}%
\providecommand \@sanitize@url [0]{\catcode `\\12\catcode `\$12\catcode
  `\&12\catcode `\#12\catcode `\^12\catcode `\_12\catcode `\%12\relax}%
\providecommand \@@startlink[1]{}%
\providecommand \@@endlink[0]{}%
\providecommand \url  [0]{\begingroup\@sanitize@url \@url }%
\providecommand \@url [1]{\endgroup\@href {#1}{\urlprefix }}%
\providecommand \urlprefix  [0]{URL }%
\providecommand \Eprint [0]{\href }%
\providecommand \doibase [0]{http://dx.doi.org/}%
\providecommand \selectlanguage [0]{\@gobble}%
\providecommand \bibinfo  [0]{\@secondoftwo}%
\providecommand \bibfield  [0]{\@secondoftwo}%
\providecommand \translation [1]{[#1]}%
\providecommand \BibitemOpen [0]{}%
\providecommand \bibitemStop [0]{}%
\providecommand \bibitemNoStop [0]{.\EOS\space}%
\providecommand \EOS [0]{\spacefactor3000\relax}%
\providecommand \BibitemShut  [1]{\csname bibitem#1\endcsname}%
\let\auto@bib@innerbib\@empty
%</preamble>
\bibitem [{\citenamefont {Guo}\ \emph {et~al.}(2024)\citenamefont {Guo},
  \citenamefont {van Delft}, \citenamefont {Gutierrez-Amigo}, \citenamefont
  {Chen}, \citenamefont {Putzke}, \citenamefont {Wagner}, \citenamefont
  {Fischer}, \citenamefont {Neupert}, \citenamefont {Errea}, \citenamefont
  {Vergniory}, \citenamefont {Wiedmann}, \citenamefont {Felser},\ and\
  \citenamefont {Moll}}]{Guo2024}%
  \BibitemOpen
  \bibfield  {author} {\bibinfo {author} {\bibfnamefont {C.}~\bibnamefont
  {Guo}}, \bibinfo {author} {\bibfnamefont {M.~R.}\ \bibnamefont {van Delft}},
  \bibinfo {author} {\bibfnamefont {M.}~\bibnamefont {Gutierrez-Amigo}},
  \bibinfo {author} {\bibfnamefont {D.}~\bibnamefont {Chen}}, \bibinfo {author}
  {\bibfnamefont {C.}~\bibnamefont {Putzke}}, \bibinfo {author} {\bibfnamefont
  {G.}~\bibnamefont {Wagner}}, \bibinfo {author} {\bibfnamefont {M.~H.}\
  \bibnamefont {Fischer}}, \bibinfo {author} {\bibfnamefont {T.}~\bibnamefont
  {Neupert}}, \bibinfo {author} {\bibfnamefont {I.}~\bibnamefont {Errea}},
  \bibinfo {author} {\bibfnamefont {M.~G.}\ \bibnamefont {Vergniory}}, \bibinfo
  {author} {\bibfnamefont {S.}~\bibnamefont {Wiedmann}}, \bibinfo {author}
  {\bibfnamefont {C.}~\bibnamefont {Felser}}, \ and\ \bibinfo {author}
  {\bibfnamefont {P.~J.~W.}\ \bibnamefont {Moll}},\ }\href {\doibase
  10.1038/s41535-024-00629-3} {\bibfield  {journal} {\bibinfo  {journal} {npj
  Quantum Materials}\ }\textbf {\bibinfo {volume} {9}},\ \bibinfo {pages} {20}
  (\bibinfo {year} {2024})}\BibitemShut {NoStop}%
\bibitem [{\citenamefont {Hu}\ \emph {et~al.}(2023)\citenamefont {Hu},
  \citenamefont {Wu}, \citenamefont {Schnyder},\ and\ \citenamefont
  {Shi}}]{Hu2023}%
  \BibitemOpen
  \bibfield  {author} {\bibinfo {author} {\bibfnamefont {Y.}~\bibnamefont
  {Hu}}, \bibinfo {author} {\bibfnamefont {X.}~\bibnamefont {Wu}}, \bibinfo
  {author} {\bibfnamefont {A.~P.}\ \bibnamefont {Schnyder}}, \ and\ \bibinfo
  {author} {\bibfnamefont {M.}~\bibnamefont {Shi}},\ }\href {\doibase
  10.1038/s41535-023-00599-y} {\bibfield  {journal} {\bibinfo  {journal} {npj
  Quantum Materials}\ }\textbf {\bibinfo {volume} {8}},\ \bibinfo {pages} {67}
  (\bibinfo {year} {2023})}\BibitemShut {NoStop}%
\bibitem [{\citenamefont {Kang}\ \emph {et~al.}(2022)\citenamefont {Kang},
  \citenamefont {Fang}, \citenamefont {Kim}, \citenamefont {Ortiz},
  \citenamefont {Ryu}, \citenamefont {Kim}, \citenamefont {Yoo}, \citenamefont
  {Sangiovanni}, \citenamefont {Di~Sante}, \citenamefont {Park}, \citenamefont
  {Jozwiak}, \citenamefont {Bostwick}, \citenamefont {Rotenberg}, \citenamefont
  {Kaxiras}, \citenamefont {Wilson}, \citenamefont {Park},\ and\ \citenamefont
  {Comin}}]{Kang2022}%
  \BibitemOpen
  \bibfield  {author} {\bibinfo {author} {\bibfnamefont {M.}~\bibnamefont
  {Kang}}, \bibinfo {author} {\bibfnamefont {S.}~\bibnamefont {Fang}}, \bibinfo
  {author} {\bibfnamefont {J.-K.}\ \bibnamefont {Kim}}, \bibinfo {author}
  {\bibfnamefont {B.~R.}\ \bibnamefont {Ortiz}}, \bibinfo {author}
  {\bibfnamefont {S.~H.}\ \bibnamefont {Ryu}}, \bibinfo {author} {\bibfnamefont
  {J.}~\bibnamefont {Kim}}, \bibinfo {author} {\bibfnamefont {J.}~\bibnamefont
  {Yoo}}, \bibinfo {author} {\bibfnamefont {G.}~\bibnamefont {Sangiovanni}},
  \bibinfo {author} {\bibfnamefont {D.}~\bibnamefont {Di~Sante}}, \bibinfo
  {author} {\bibfnamefont {B.-G.}\ \bibnamefont {Park}}, \bibinfo {author}
  {\bibfnamefont {C.}~\bibnamefont {Jozwiak}}, \bibinfo {author} {\bibfnamefont
  {A.}~\bibnamefont {Bostwick}}, \bibinfo {author} {\bibfnamefont
  {E.}~\bibnamefont {Rotenberg}}, \bibinfo {author} {\bibfnamefont
  {E.}~\bibnamefont {Kaxiras}}, \bibinfo {author} {\bibfnamefont {S.~D.}\
  \bibnamefont {Wilson}}, \bibinfo {author} {\bibfnamefont {J.-H.}\
  \bibnamefont {Park}}, \ and\ \bibinfo {author} {\bibfnamefont
  {R.}~\bibnamefont {Comin}},\ }\href {\doibase 10.1038/s41567-021-01451-5}
  {\bibfield  {journal} {\bibinfo  {journal} {Nature Physics}\ }\textbf
  {\bibinfo {volume} {18}},\ \bibinfo {pages} {301} (\bibinfo {year}
  {2022})}\BibitemShut {NoStop}%
\bibitem [{\citenamefont {Sato}\ and\ \citenamefont {Ando}(2017)}]{Sato2017}%
  \BibitemOpen
  \bibfield  {author} {\bibinfo {author} {\bibfnamefont {M.}~\bibnamefont
  {Sato}}\ and\ \bibinfo {author} {\bibfnamefont {Y.}~\bibnamefont {Ando}},\
  }\href {\doibase 10.1088/1361-6633/aa6ac7} {\bibfield  {journal} {\bibinfo
  {journal} {Reports on Progress in Physics}\ }\textbf {\bibinfo {volume}
  {80}},\ \bibinfo {pages} {076501} (\bibinfo {year} {2017})}\BibitemShut
  {NoStop}%
\bibitem [{\citenamefont {Xing}\ \emph {et~al.}(2024)\citenamefont {Xing},
  \citenamefont {Bae}, \citenamefont {Ritz}, \citenamefont {Yang},
  \citenamefont {Birol}, \citenamefont {Capa~Salinas}, \citenamefont {Ortiz},
  \citenamefont {Wilson}, \citenamefont {Wang}, \citenamefont {Fernandes},\
  and\ \citenamefont {Madhavan}}]{Xing2024}%
  \BibitemOpen
  \bibfield  {author} {\bibinfo {author} {\bibfnamefont {Y.}~\bibnamefont
  {Xing}}, \bibinfo {author} {\bibfnamefont {S.}~\bibnamefont {Bae}}, \bibinfo
  {author} {\bibfnamefont {E.}~\bibnamefont {Ritz}}, \bibinfo {author}
  {\bibfnamefont {F.}~\bibnamefont {Yang}}, \bibinfo {author} {\bibfnamefont
  {T.}~\bibnamefont {Birol}}, \bibinfo {author} {\bibfnamefont {A.~N.}\
  \bibnamefont {Capa~Salinas}}, \bibinfo {author} {\bibfnamefont {B.~R.}\
  \bibnamefont {Ortiz}}, \bibinfo {author} {\bibfnamefont {S.~D.}\ \bibnamefont
  {Wilson}}, \bibinfo {author} {\bibfnamefont {Z.}~\bibnamefont {Wang}},
  \bibinfo {author} {\bibfnamefont {R.~M.}\ \bibnamefont {Fernandes}}, \ and\
  \bibinfo {author} {\bibfnamefont {V.}~\bibnamefont {Madhavan}},\ }\href
  {\doibase 10.1038/s41586-024-07519-5} {\bibfield  {journal} {\bibinfo
  {journal} {Nature}\ }\textbf {\bibinfo {volume} {631}},\ \bibinfo {pages}
  {60} (\bibinfo {year} {2024})}\BibitemShut {NoStop}%
\bibitem [{\citenamefont {Xu}\ \emph {et~al.}(2015)\citenamefont {Xu},
  \citenamefont {Lian},\ and\ \citenamefont {Zhang}}]{Xu2015}%
  \BibitemOpen
  \bibfield  {author} {\bibinfo {author} {\bibfnamefont {G.}~\bibnamefont
  {Xu}}, \bibinfo {author} {\bibfnamefont {B.}~\bibnamefont {Lian}}, \ and\
  \bibinfo {author} {\bibfnamefont {S.-C.}\ \bibnamefont {Zhang}},\ }\href
  {\doibase 10.1103/physrevlett.115.186802} {\bibfield  {journal} {\bibinfo
  {journal} {Physical Review Letters}\ }\textbf {\bibinfo {volume} {115}},\
  \bibinfo {pages} {186802} (\bibinfo {year} {2015})}\BibitemShut {NoStop}%
\bibitem [{\citenamefont {Guo}\ and\ \citenamefont {Franz}(2009)}]{Guo2009}%
  \BibitemOpen
  \bibfield  {author} {\bibinfo {author} {\bibfnamefont {H.-M.}\ \bibnamefont
  {Guo}}\ and\ \bibinfo {author} {\bibfnamefont {M.}~\bibnamefont {Franz}},\
  }\href {\doibase 10.1103/physrevb.80.113102} {\bibfield  {journal} {\bibinfo
  {journal} {Physical Review B}\ }\textbf {\bibinfo {volume} {80}},\ \bibinfo
  {pages} {113102} (\bibinfo {year} {2009})}\BibitemShut {NoStop}%
\bibitem [{\citenamefont {Tang}\ \emph {et~al.}(2011)\citenamefont {Tang},
  \citenamefont {Mei},\ and\ \citenamefont {Wen}}]{Tang2011}%
  \BibitemOpen
  \bibfield  {author} {\bibinfo {author} {\bibfnamefont {E.}~\bibnamefont
  {Tang}}, \bibinfo {author} {\bibfnamefont {J.-W.}\ \bibnamefont {Mei}}, \
  and\ \bibinfo {author} {\bibfnamefont {X.-G.}\ \bibnamefont {Wen}},\ }\href
  {\doibase 10.1103/physrevlett.106.236802} {\bibfield  {journal} {\bibinfo
  {journal} {Physical Review Letters}\ }\textbf {\bibinfo {volume} {106}},\
  \bibinfo {pages} {236802} (\bibinfo {year} {2011})}\BibitemShut {NoStop}%
\bibitem [{\citenamefont {Wang}\ \emph {et~al.}(2013)\citenamefont {Wang},
  \citenamefont {Li}, \citenamefont {Xiang},\ and\ \citenamefont
  {Wang}}]{Wang2013}%
  \BibitemOpen
  \bibfield  {author} {\bibinfo {author} {\bibfnamefont {W.-S.}\ \bibnamefont
  {Wang}}, \bibinfo {author} {\bibfnamefont {Z.-Z.}\ \bibnamefont {Li}},
  \bibinfo {author} {\bibfnamefont {Y.-Y.}\ \bibnamefont {Xiang}}, \ and\
  \bibinfo {author} {\bibfnamefont {Q.-H.}\ \bibnamefont {Wang}},\ }\href
  {\doibase 10.1103/physrevb.87.115135} {\bibfield  {journal} {\bibinfo
  {journal} {Physical Review B}\ }\textbf {\bibinfo {volume} {87}},\ \bibinfo
  {pages} {115135} (\bibinfo {year} {2013})}\BibitemShut {NoStop}%
\bibitem [{\citenamefont {Kiesel}\ \emph {et~al.}(2013)\citenamefont {Kiesel},
  \citenamefont {Platt},\ and\ \citenamefont {Thomale}}]{Kiesel2013}%
  \BibitemOpen
  \bibfield  {author} {\bibinfo {author} {\bibfnamefont {M.~L.}\ \bibnamefont
  {Kiesel}}, \bibinfo {author} {\bibfnamefont {C.}~\bibnamefont {Platt}}, \
  and\ \bibinfo {author} {\bibfnamefont {R.}~\bibnamefont {Thomale}},\ }\href
  {\doibase 10.1103/physrevlett.110.126405} {\bibfield  {journal} {\bibinfo
  {journal} {Physical Review Letters}\ }\textbf {\bibinfo {volume} {110}},\
  \bibinfo {pages} {126405} (\bibinfo {year} {2013})}\BibitemShut {NoStop}%
\bibitem [{\citenamefont {Kiesel}\ and\ \citenamefont
  {Thomale}(2012)}]{Kiesel2012}%
  \BibitemOpen
  \bibfield  {author} {\bibinfo {author} {\bibfnamefont {M.~L.}\ \bibnamefont
  {Kiesel}}\ and\ \bibinfo {author} {\bibfnamefont {R.}~\bibnamefont
  {Thomale}},\ }\href {\doibase 10.1103/physrevb.86.121105} {\bibfield
  {journal} {\bibinfo  {journal} {Physical Review B}\ }\textbf {\bibinfo
  {volume} {86}},\ \bibinfo {pages} {121105(R)} (\bibinfo {year}
  {2012})}\BibitemShut {NoStop}%
\bibitem [{\citenamefont {Yu}\ and\ \citenamefont {Li}(2012)}]{Yu2012}%
  \BibitemOpen
  \bibfield  {author} {\bibinfo {author} {\bibfnamefont {S.-L.}\ \bibnamefont
  {Yu}}\ and\ \bibinfo {author} {\bibfnamefont {J.-X.}\ \bibnamefont {Li}},\
  }\href {\doibase 10.1103/physrevb.85.144402} {\bibfield  {journal} {\bibinfo
  {journal} {Physical Review B}\ }\textbf {\bibinfo {volume} {85}},\ \bibinfo
  {pages} {144402} (\bibinfo {year} {2012})}\BibitemShut {NoStop}%
\bibitem [{\citenamefont {Mielke}(1992)}]{Mielke1992}%
  \BibitemOpen
  \bibfield  {author} {\bibinfo {author} {\bibfnamefont {A.}~\bibnamefont
  {Mielke}},\ }\href {\doibase 10.1088/0305-4470/25/16/011} {\bibfield
  {journal} {\bibinfo  {journal} {Journal of Physics A: Mathematical and
  General}\ }\textbf {\bibinfo {volume} {25}},\ \bibinfo {pages} {4335}
  (\bibinfo {year} {1992})}\BibitemShut {NoStop}%
\bibitem [{\citenamefont {Vonsovsky}\ and\ \citenamefont
  {Katsnelson}(1989)}]{Vonsovsky1989}%
  \BibitemOpen
  \bibfield  {author} {\bibinfo {author} {\bibfnamefont {S.}~\bibnamefont
  {Vonsovsky}}\ and\ \bibinfo {author} {\bibfnamefont {M.}~\bibnamefont
  {Katsnelson}},\ }\href {\doibase 10.1016/s0921-4526(89)80054-7} {\bibfield
  {journal} {\bibinfo  {journal} {Physica B: Condensed Matter}\ }\textbf
  {\bibinfo {volume} {159}},\ \bibinfo {pages} {61} (\bibinfo {year}
  {1989})}\BibitemShut {NoStop}%
\bibitem [{\citenamefont {Frachet}\ \emph {et~al.}(2024)\citenamefont
  {Frachet}, \citenamefont {Wang}, \citenamefont {Xia}, \citenamefont {Guo},
  \citenamefont {He}, \citenamefont {Maraytta}, \citenamefont {Heid},
  \citenamefont {Haghighirad}, \citenamefont {Merz}, \citenamefont {Meingast},\
  and\ \citenamefont {Hardy}}]{Frachet2024}%
  \BibitemOpen
  \bibfield  {author} {\bibinfo {author} {\bibfnamefont {M.}~\bibnamefont
  {Frachet}}, \bibinfo {author} {\bibfnamefont {L.}~\bibnamefont {Wang}},
  \bibinfo {author} {\bibfnamefont {W.}~\bibnamefont {Xia}}, \bibinfo {author}
  {\bibfnamefont {Y.}~\bibnamefont {Guo}}, \bibinfo {author} {\bibfnamefont
  {M.}~\bibnamefont {He}}, \bibinfo {author} {\bibfnamefont {N.}~\bibnamefont
  {Maraytta}}, \bibinfo {author} {\bibfnamefont {R.}~\bibnamefont {Heid}},
  \bibinfo {author} {\bibfnamefont {A.-A.}\ \bibnamefont {Haghighirad}},
  \bibinfo {author} {\bibfnamefont {M.}~\bibnamefont {Merz}}, \bibinfo {author}
  {\bibfnamefont {C.}~\bibnamefont {Meingast}}, \ and\ \bibinfo {author}
  {\bibfnamefont {F.}~\bibnamefont {Hardy}},\ }\href {\doibase
  10.1103/physrevlett.132.186001} {\bibfield  {journal} {\bibinfo  {journal}
  {Physical Review Letters}\ }\textbf {\bibinfo {volume} {132}},\ \bibinfo
  {pages} {186001} (\bibinfo {year} {2024})}\BibitemShut {NoStop}%
\bibitem [{\citenamefont {Xu}\ \emph {et~al.}(2022)\citenamefont {Xu},
  \citenamefont {Ni}, \citenamefont {Liu}, \citenamefont {Ortiz}, \citenamefont
  {Deng}, \citenamefont {Wilson}, \citenamefont {Yan}, \citenamefont
  {Balents},\ and\ \citenamefont {Wu}}]{Xu2022}%
  \BibitemOpen
  \bibfield  {author} {\bibinfo {author} {\bibfnamefont {Y.}~\bibnamefont
  {Xu}}, \bibinfo {author} {\bibfnamefont {Z.}~\bibnamefont {Ni}}, \bibinfo
  {author} {\bibfnamefont {Y.}~\bibnamefont {Liu}}, \bibinfo {author}
  {\bibfnamefont {B.~R.}\ \bibnamefont {Ortiz}}, \bibinfo {author}
  {\bibfnamefont {Q.}~\bibnamefont {Deng}}, \bibinfo {author} {\bibfnamefont
  {S.~D.}\ \bibnamefont {Wilson}}, \bibinfo {author} {\bibfnamefont
  {B.}~\bibnamefont {Yan}}, \bibinfo {author} {\bibfnamefont {L.}~\bibnamefont
  {Balents}}, \ and\ \bibinfo {author} {\bibfnamefont {L.}~\bibnamefont {Wu}},\
  }\href {\doibase 10.1038/s41567-022-01805-7} {\bibfield  {journal} {\bibinfo
  {journal} {Nature Physics}\ }\textbf {\bibinfo {volume} {18}},\ \bibinfo
  {pages} {1470} (\bibinfo {year} {2022})}\BibitemShut {NoStop}%
\bibitem [{\citenamefont {Nie}\ \emph {et~al.}(2022)\citenamefont {Nie},
  \citenamefont {Sun}, \citenamefont {Ma}, \citenamefont {Song}, \citenamefont
  {Zheng}, \citenamefont {Liang}, \citenamefont {Wu}, \citenamefont {Yu},
  \citenamefont {Li}, \citenamefont {Shan}, \citenamefont {Zhao}, \citenamefont
  {Li}, \citenamefont {Kang}, \citenamefont {Wu}, \citenamefont {Zhou},
  \citenamefont {Liu}, \citenamefont {Xiang}, \citenamefont {Ying},
  \citenamefont {Wang}, \citenamefont {Wu},\ and\ \citenamefont
  {Chen}}]{Nie2022}%
  \BibitemOpen
  \bibfield  {author} {\bibinfo {author} {\bibfnamefont {L.}~\bibnamefont
  {Nie}}, \bibinfo {author} {\bibfnamefont {K.}~\bibnamefont {Sun}}, \bibinfo
  {author} {\bibfnamefont {W.}~\bibnamefont {Ma}}, \bibinfo {author}
  {\bibfnamefont {D.}~\bibnamefont {Song}}, \bibinfo {author} {\bibfnamefont
  {L.}~\bibnamefont {Zheng}}, \bibinfo {author} {\bibfnamefont
  {Z.}~\bibnamefont {Liang}}, \bibinfo {author} {\bibfnamefont
  {P.}~\bibnamefont {Wu}}, \bibinfo {author} {\bibfnamefont {F.}~\bibnamefont
  {Yu}}, \bibinfo {author} {\bibfnamefont {J.}~\bibnamefont {Li}}, \bibinfo
  {author} {\bibfnamefont {M.}~\bibnamefont {Shan}}, \bibinfo {author}
  {\bibfnamefont {D.}~\bibnamefont {Zhao}}, \bibinfo {author} {\bibfnamefont
  {S.}~\bibnamefont {Li}}, \bibinfo {author} {\bibfnamefont {B.}~\bibnamefont
  {Kang}}, \bibinfo {author} {\bibfnamefont {Z.}~\bibnamefont {Wu}}, \bibinfo
  {author} {\bibfnamefont {Y.}~\bibnamefont {Zhou}}, \bibinfo {author}
  {\bibfnamefont {K.}~\bibnamefont {Liu}}, \bibinfo {author} {\bibfnamefont
  {Z.}~\bibnamefont {Xiang}}, \bibinfo {author} {\bibfnamefont
  {J.}~\bibnamefont {Ying}}, \bibinfo {author} {\bibfnamefont {Z.}~\bibnamefont
  {Wang}}, \bibinfo {author} {\bibfnamefont {T.}~\bibnamefont {Wu}}, \ and\
  \bibinfo {author} {\bibfnamefont {X.}~\bibnamefont {Chen}},\ }\href {\doibase
  10.1038/s41586-022-04493-8} {\bibfield  {journal} {\bibinfo  {journal}
  {Nature}\ }\textbf {\bibinfo {volume} {604}},\ \bibinfo {pages} {59}
  (\bibinfo {year} {2022})}\BibitemShut {NoStop}%
\bibitem [{\citenamefont {Zhao}\ \emph {et~al.}(2021)\citenamefont {Zhao},
  \citenamefont {Li}, \citenamefont {Ortiz}, \citenamefont {Teicher},
  \citenamefont {Park}, \citenamefont {Ye}, \citenamefont {Wang}, \citenamefont
  {Balents}, \citenamefont {Wilson},\ and\ \citenamefont
  {Zeljkovic}}]{Zhao2021}%
  \BibitemOpen
  \bibfield  {author} {\bibinfo {author} {\bibfnamefont {H.}~\bibnamefont
  {Zhao}}, \bibinfo {author} {\bibfnamefont {H.}~\bibnamefont {Li}}, \bibinfo
  {author} {\bibfnamefont {B.~R.}\ \bibnamefont {Ortiz}}, \bibinfo {author}
  {\bibfnamefont {S.~M.~L.}\ \bibnamefont {Teicher}}, \bibinfo {author}
  {\bibfnamefont {T.}~\bibnamefont {Park}}, \bibinfo {author} {\bibfnamefont
  {M.}~\bibnamefont {Ye}}, \bibinfo {author} {\bibfnamefont {Z.}~\bibnamefont
  {Wang}}, \bibinfo {author} {\bibfnamefont {L.}~\bibnamefont {Balents}},
  \bibinfo {author} {\bibfnamefont {S.~D.}\ \bibnamefont {Wilson}}, \ and\
  \bibinfo {author} {\bibfnamefont {I.}~\bibnamefont {Zeljkovic}},\ }\href
  {\doibase 10.1038/s41586-021-03946-w} {\bibfield  {journal} {\bibinfo
  {journal} {Nature}\ }\textbf {\bibinfo {volume} {599}},\ \bibinfo {pages}
  {216} (\bibinfo {year} {2021})}\BibitemShut {NoStop}%
\bibitem [{\citenamefont {Li}\ \emph {et~al.}(2021)\citenamefont {Li},
  \citenamefont {Zhang}, \citenamefont {Yilmaz}, \citenamefont {Pai},
  \citenamefont {Marvinney}, \citenamefont {Said}, \citenamefont {Yin},
  \citenamefont {Gong}, \citenamefont {Tu}, \citenamefont {Vescovo},
  \citenamefont {Nelson}, \citenamefont {Moore}, \citenamefont {Murakami},
  \citenamefont {Lei}, \citenamefont {Lee}, \citenamefont {Lawrie},\ and\
  \citenamefont {Miao}}]{Li2021}%
  \BibitemOpen
  \bibfield  {author} {\bibinfo {author} {\bibfnamefont {H.}~\bibnamefont
  {Li}}, \bibinfo {author} {\bibfnamefont {T.}~\bibnamefont {Zhang}}, \bibinfo
  {author} {\bibfnamefont {T.}~\bibnamefont {Yilmaz}}, \bibinfo {author}
  {\bibfnamefont {Y.}~\bibnamefont {Pai}}, \bibinfo {author} {\bibfnamefont
  {C.}~\bibnamefont {Marvinney}}, \bibinfo {author} {\bibfnamefont
  {A.}~\bibnamefont {Said}}, \bibinfo {author} {\bibfnamefont {Q.}~\bibnamefont
  {Yin}}, \bibinfo {author} {\bibfnamefont {C.}~\bibnamefont {Gong}}, \bibinfo
  {author} {\bibfnamefont {Z.}~\bibnamefont {Tu}}, \bibinfo {author}
  {\bibfnamefont {E.}~\bibnamefont {Vescovo}}, \bibinfo {author} {\bibfnamefont
  {C.}~\bibnamefont {Nelson}}, \bibinfo {author} {\bibfnamefont
  {R.}~\bibnamefont {Moore}}, \bibinfo {author} {\bibfnamefont
  {S.}~\bibnamefont {Murakami}}, \bibinfo {author} {\bibfnamefont
  {H.}~\bibnamefont {Lei}}, \bibinfo {author} {\bibfnamefont {H.}~\bibnamefont
  {Lee}}, \bibinfo {author} {\bibfnamefont {B.}~\bibnamefont {Lawrie}}, \ and\
  \bibinfo {author} {\bibfnamefont {H.}~\bibnamefont {Miao}},\ }\href {\doibase
  10.1103/physrevx.11.031050} {\bibfield  {journal} {\bibinfo  {journal}
  {Physical Review X}\ }\textbf {\bibinfo {volume} {11}},\ \bibinfo {pages}
  {031050} (\bibinfo {year} {2021})}\BibitemShut {NoStop}%
\bibitem [{\citenamefont {Liang}\ \emph {et~al.}(2021)\citenamefont {Liang},
  \citenamefont {Hou}, \citenamefont {Zhang}, \citenamefont {Ma}, \citenamefont
  {Wu}, \citenamefont {Zhang}, \citenamefont {Yu}, \citenamefont {Ying},
  \citenamefont {Jiang}, \citenamefont {Shan}, \citenamefont {Wang},\ and\
  \citenamefont {Chen}}]{Liang2021}%
  \BibitemOpen
  \bibfield  {author} {\bibinfo {author} {\bibfnamefont {Z.}~\bibnamefont
  {Liang}}, \bibinfo {author} {\bibfnamefont {X.}~\bibnamefont {Hou}}, \bibinfo
  {author} {\bibfnamefont {F.}~\bibnamefont {Zhang}}, \bibinfo {author}
  {\bibfnamefont {W.}~\bibnamefont {Ma}}, \bibinfo {author} {\bibfnamefont
  {P.}~\bibnamefont {Wu}}, \bibinfo {author} {\bibfnamefont {Z.}~\bibnamefont
  {Zhang}}, \bibinfo {author} {\bibfnamefont {F.}~\bibnamefont {Yu}}, \bibinfo
  {author} {\bibfnamefont {J.-J.}\ \bibnamefont {Ying}}, \bibinfo {author}
  {\bibfnamefont {K.}~\bibnamefont {Jiang}}, \bibinfo {author} {\bibfnamefont
  {L.}~\bibnamefont {Shan}}, \bibinfo {author} {\bibfnamefont {Z.}~\bibnamefont
  {Wang}}, \ and\ \bibinfo {author} {\bibfnamefont {X.-H.}\ \bibnamefont
  {Chen}},\ }\href {\doibase 10.1103/physrevx.11.031026} {\bibfield  {journal}
  {\bibinfo  {journal} {Physical Review X}\ }\textbf {\bibinfo {volume} {11}},\
  \bibinfo {pages} {031026} (\bibinfo {year} {2021})}\BibitemShut {NoStop}%
\bibitem [{\citenamefont {Chen}\ \emph
  {et~al.}(2021{\natexlab{a}})\citenamefont {Chen}, \citenamefont {Yang},
  \citenamefont {Hu}, \citenamefont {Zhao}, \citenamefont {Yuan}, \citenamefont
  {Xing}, \citenamefont {Qian}, \citenamefont {Huang}, \citenamefont {Li},
  \citenamefont {Ye}, \citenamefont {Ma}, \citenamefont {Ni}, \citenamefont
  {Zhang}, \citenamefont {Yin}, \citenamefont {Gong}, \citenamefont {Tu},
  \citenamefont {Lei}, \citenamefont {Tan}, \citenamefont {Zhou}, \citenamefont
  {Shen}, \citenamefont {Dong}, \citenamefont {Yan}, \citenamefont {Wang},\
  and\ \citenamefont {Gao}}]{Chen2021}%
  \BibitemOpen
  \bibfield  {author} {\bibinfo {author} {\bibfnamefont {H.}~\bibnamefont
  {Chen}}, \bibinfo {author} {\bibfnamefont {H.}~\bibnamefont {Yang}}, \bibinfo
  {author} {\bibfnamefont {B.}~\bibnamefont {Hu}}, \bibinfo {author}
  {\bibfnamefont {Z.}~\bibnamefont {Zhao}}, \bibinfo {author} {\bibfnamefont
  {J.}~\bibnamefont {Yuan}}, \bibinfo {author} {\bibfnamefont {Y.}~\bibnamefont
  {Xing}}, \bibinfo {author} {\bibfnamefont {G.}~\bibnamefont {Qian}}, \bibinfo
  {author} {\bibfnamefont {Z.}~\bibnamefont {Huang}}, \bibinfo {author}
  {\bibfnamefont {G.}~\bibnamefont {Li}}, \bibinfo {author} {\bibfnamefont
  {Y.}~\bibnamefont {Ye}}, \bibinfo {author} {\bibfnamefont {S.}~\bibnamefont
  {Ma}}, \bibinfo {author} {\bibfnamefont {S.}~\bibnamefont {Ni}}, \bibinfo
  {author} {\bibfnamefont {H.}~\bibnamefont {Zhang}}, \bibinfo {author}
  {\bibfnamefont {Q.}~\bibnamefont {Yin}}, \bibinfo {author} {\bibfnamefont
  {C.}~\bibnamefont {Gong}}, \bibinfo {author} {\bibfnamefont {Z.}~\bibnamefont
  {Tu}}, \bibinfo {author} {\bibfnamefont {H.}~\bibnamefont {Lei}}, \bibinfo
  {author} {\bibfnamefont {H.}~\bibnamefont {Tan}}, \bibinfo {author}
  {\bibfnamefont {S.}~\bibnamefont {Zhou}}, \bibinfo {author} {\bibfnamefont
  {C.}~\bibnamefont {Shen}}, \bibinfo {author} {\bibfnamefont {X.}~\bibnamefont
  {Dong}}, \bibinfo {author} {\bibfnamefont {B.}~\bibnamefont {Yan}}, \bibinfo
  {author} {\bibfnamefont {Z.}~\bibnamefont {Wang}}, \ and\ \bibinfo {author}
  {\bibfnamefont {H.-J.}\ \bibnamefont {Gao}},\ }\href {\doibase
  10.1038/s41586-021-03983-5} {\bibfield  {journal} {\bibinfo  {journal}
  {Nature}\ }\textbf {\bibinfo {volume} {599}},\ \bibinfo {pages} {222}
  (\bibinfo {year} {2021}{\natexlab{a}})}\BibitemShut {NoStop}%
\bibitem [{\citenamefont {Guguchia}\ \emph {et~al.}(2023)\citenamefont
  {Guguchia}, \citenamefont {Mielke}, \citenamefont {Das}, \citenamefont
  {Gupta}, \citenamefont {Yin}, \citenamefont {Liu}, \citenamefont {Yin},
  \citenamefont {Christensen}, \citenamefont {Tu}, \citenamefont {Gong},
  \citenamefont {Shumiya}, \citenamefont {Hossain}, \citenamefont
  {Gamsakhurdashvili}, \citenamefont {Elender}, \citenamefont {Dai},
  \citenamefont {Amato}, \citenamefont {Shi}, \citenamefont {Lei},
  \citenamefont {Fernandes}, \citenamefont {Hasan}, \citenamefont {Luetkens},\
  and\ \citenamefont {Khasanov}}]{Guguchia2023}%
  \BibitemOpen
  \bibfield  {author} {\bibinfo {author} {\bibfnamefont {Z.}~\bibnamefont
  {Guguchia}}, \bibinfo {author} {\bibfnamefont {C.}~\bibnamefont {Mielke}},
  \bibinfo {author} {\bibfnamefont {D.}~\bibnamefont {Das}}, \bibinfo {author}
  {\bibfnamefont {R.}~\bibnamefont {Gupta}}, \bibinfo {author} {\bibfnamefont
  {J.-X.}\ \bibnamefont {Yin}}, \bibinfo {author} {\bibfnamefont
  {H.}~\bibnamefont {Liu}}, \bibinfo {author} {\bibfnamefont {Q.}~\bibnamefont
  {Yin}}, \bibinfo {author} {\bibfnamefont {M.~H.}\ \bibnamefont
  {Christensen}}, \bibinfo {author} {\bibfnamefont {Z.}~\bibnamefont {Tu}},
  \bibinfo {author} {\bibfnamefont {C.}~\bibnamefont {Gong}}, \bibinfo {author}
  {\bibfnamefont {N.}~\bibnamefont {Shumiya}}, \bibinfo {author} {\bibfnamefont
  {M.~S.}\ \bibnamefont {Hossain}}, \bibinfo {author} {\bibfnamefont
  {T.}~\bibnamefont {Gamsakhurdashvili}}, \bibinfo {author} {\bibfnamefont
  {M.}~\bibnamefont {Elender}}, \bibinfo {author} {\bibfnamefont
  {P.}~\bibnamefont {Dai}}, \bibinfo {author} {\bibfnamefont {A.}~\bibnamefont
  {Amato}}, \bibinfo {author} {\bibfnamefont {Y.}~\bibnamefont {Shi}}, \bibinfo
  {author} {\bibfnamefont {H.~C.}\ \bibnamefont {Lei}}, \bibinfo {author}
  {\bibfnamefont {R.~M.}\ \bibnamefont {Fernandes}}, \bibinfo {author}
  {\bibfnamefont {M.~Z.}\ \bibnamefont {Hasan}}, \bibinfo {author}
  {\bibfnamefont {H.}~\bibnamefont {Luetkens}}, \ and\ \bibinfo {author}
  {\bibfnamefont {R.}~\bibnamefont {Khasanov}},\ }\href {\doibase
  10.1038/s41467-022-35718-z} {\bibfield  {journal} {\bibinfo  {journal}
  {Nature Communications}\ }\textbf {\bibinfo {volume} {14}},\ \bibinfo {pages}
  {153} (\bibinfo {year} {2023})}\BibitemShut {NoStop}%
\bibitem [{\citenamefont {Jiang}\ \emph {et~al.}(2022)\citenamefont {Jiang},
  \citenamefont {Wu}, \citenamefont {Yin}, \citenamefont {Wang}, \citenamefont
  {Hasan}, \citenamefont {Wilson}, \citenamefont {Chen},\ and\ \citenamefont
  {Hu}}]{Jiang2022}%
  \BibitemOpen
  \bibfield  {author} {\bibinfo {author} {\bibfnamefont {K.}~\bibnamefont
  {Jiang}}, \bibinfo {author} {\bibfnamefont {T.}~\bibnamefont {Wu}}, \bibinfo
  {author} {\bibfnamefont {J.-X.}\ \bibnamefont {Yin}}, \bibinfo {author}
  {\bibfnamefont {Z.}~\bibnamefont {Wang}}, \bibinfo {author} {\bibfnamefont
  {M.~Z.}\ \bibnamefont {Hasan}}, \bibinfo {author} {\bibfnamefont {S.~D.}\
  \bibnamefont {Wilson}}, \bibinfo {author} {\bibfnamefont {X.}~\bibnamefont
  {Chen}}, \ and\ \bibinfo {author} {\bibfnamefont {J.}~\bibnamefont {Hu}},\
  }\href {\doibase 10.1093/nsr/nwac199} {\bibfield  {journal} {\bibinfo
  {journal} {National Science Review}\ }\textbf {\bibinfo {volume} {10}},\
  \bibinfo {pages} {nwac199} (\bibinfo {year} {2022})}\BibitemShut {NoStop}%
\bibitem [{\citenamefont {Zhang}\ \emph {et~al.}(2021)\citenamefont {Zhang},
  \citenamefont {Chen}, \citenamefont {Zhou}, \citenamefont {Yuan},
  \citenamefont {Wang}, \citenamefont {Wang}, \citenamefont {Yang},
  \citenamefont {An}, \citenamefont {Zhang}, \citenamefont {Zhu}, \citenamefont
  {Zhou}, \citenamefont {Chen}, \citenamefont {Zhou},\ and\ \citenamefont
  {Yang}}]{Zhang2021}%
  \BibitemOpen
  \bibfield  {author} {\bibinfo {author} {\bibfnamefont {Z.}~\bibnamefont
  {Zhang}}, \bibinfo {author} {\bibfnamefont {Z.}~\bibnamefont {Chen}},
  \bibinfo {author} {\bibfnamefont {Y.}~\bibnamefont {Zhou}}, \bibinfo {author}
  {\bibfnamefont {Y.}~\bibnamefont {Yuan}}, \bibinfo {author} {\bibfnamefont
  {S.}~\bibnamefont {Wang}}, \bibinfo {author} {\bibfnamefont {J.}~\bibnamefont
  {Wang}}, \bibinfo {author} {\bibfnamefont {H.}~\bibnamefont {Yang}}, \bibinfo
  {author} {\bibfnamefont {C.}~\bibnamefont {An}}, \bibinfo {author}
  {\bibfnamefont {L.}~\bibnamefont {Zhang}}, \bibinfo {author} {\bibfnamefont
  {X.}~\bibnamefont {Zhu}}, \bibinfo {author} {\bibfnamefont {Y.}~\bibnamefont
  {Zhou}}, \bibinfo {author} {\bibfnamefont {X.}~\bibnamefont {Chen}}, \bibinfo
  {author} {\bibfnamefont {J.}~\bibnamefont {Zhou}}, \ and\ \bibinfo {author}
  {\bibfnamefont {Z.}~\bibnamefont {Yang}},\ }\href {\doibase
  10.1103/physrevb.103.224513} {\bibfield  {journal} {\bibinfo  {journal}
  {Physical Review B}\ }\textbf {\bibinfo {volume} {103}},\ \bibinfo {pages}
  {224513} (\bibinfo {year} {2021})}\BibitemShut {NoStop}%
\bibitem [{\citenamefont {Neupert}\ \emph {et~al.}(2021)\citenamefont
  {Neupert}, \citenamefont {Denner}, \citenamefont {Yin}, \citenamefont
  {Thomale},\ and\ \citenamefont {Hasan}}]{Neupert2021}%
  \BibitemOpen
  \bibfield  {author} {\bibinfo {author} {\bibfnamefont {T.}~\bibnamefont
  {Neupert}}, \bibinfo {author} {\bibfnamefont {M.~M.}\ \bibnamefont {Denner}},
  \bibinfo {author} {\bibfnamefont {J.-X.}\ \bibnamefont {Yin}}, \bibinfo
  {author} {\bibfnamefont {R.}~\bibnamefont {Thomale}}, \ and\ \bibinfo
  {author} {\bibfnamefont {M.~Z.}\ \bibnamefont {Hasan}},\ }\href {\doibase
  10.1038/s41567-021-01404-y} {\bibfield  {journal} {\bibinfo  {journal}
  {Nature Physics}\ }\textbf {\bibinfo {volume} {18}},\ \bibinfo {pages} {137}
  (\bibinfo {year} {2021})}\BibitemShut {NoStop}%
\bibitem [{\citenamefont {Chen}\ \emph
  {et~al.}(2021{\natexlab{b}})\citenamefont {Chen}, \citenamefont {Wang},
  \citenamefont {Yin}, \citenamefont {Gu}, \citenamefont {Jiang}, \citenamefont
  {Tu}, \citenamefont {Gong}, \citenamefont {Uwatoko}, \citenamefont {Sun},
  \citenamefont {Lei}, \citenamefont {Hu},\ and\ \citenamefont
  {Cheng}}]{Chen2021a}%
  \BibitemOpen
  \bibfield  {author} {\bibinfo {author} {\bibfnamefont {K.}~\bibnamefont
  {Chen}}, \bibinfo {author} {\bibfnamefont {N.}~\bibnamefont {Wang}}, \bibinfo
  {author} {\bibfnamefont {Q.}~\bibnamefont {Yin}}, \bibinfo {author}
  {\bibfnamefont {Y.}~\bibnamefont {Gu}}, \bibinfo {author} {\bibfnamefont
  {K.}~\bibnamefont {Jiang}}, \bibinfo {author} {\bibfnamefont
  {Z.}~\bibnamefont {Tu}}, \bibinfo {author} {\bibfnamefont {C.}~\bibnamefont
  {Gong}}, \bibinfo {author} {\bibfnamefont {Y.}~\bibnamefont {Uwatoko}},
  \bibinfo {author} {\bibfnamefont {J.}~\bibnamefont {Sun}}, \bibinfo {author}
  {\bibfnamefont {H.}~\bibnamefont {Lei}}, \bibinfo {author} {\bibfnamefont
  {J.}~\bibnamefont {Hu}}, \ and\ \bibinfo {author} {\bibfnamefont {J.-G.}\
  \bibnamefont {Cheng}},\ }\href {\doibase 10.1103/physrevlett.126.247001}
  {\bibfield  {journal} {\bibinfo  {journal} {Physical Review Letters}\
  }\textbf {\bibinfo {volume} {126}},\ \bibinfo {pages} {247001} (\bibinfo
  {year} {2021}{\natexlab{b}})}\BibitemShut {NoStop}%
\bibitem [{\citenamefont {Chen}\ \emph
  {et~al.}(2021{\natexlab{c}})\citenamefont {Chen}, \citenamefont {Zhan},
  \citenamefont {Wang}, \citenamefont {Deng}, \citenamefont {Liu},
  \citenamefont {Chen}, \citenamefont {Guo},\ and\ \citenamefont
  {Chen}}]{Chen2021b}%
  \BibitemOpen
  \bibfield  {author} {\bibinfo {author} {\bibfnamefont {X.}~\bibnamefont
  {Chen}}, \bibinfo {author} {\bibfnamefont {X.}~\bibnamefont {Zhan}}, \bibinfo
  {author} {\bibfnamefont {X.}~\bibnamefont {Wang}}, \bibinfo {author}
  {\bibfnamefont {J.}~\bibnamefont {Deng}}, \bibinfo {author} {\bibfnamefont
  {X.-B.}\ \bibnamefont {Liu}}, \bibinfo {author} {\bibfnamefont
  {X.}~\bibnamefont {Chen}}, \bibinfo {author} {\bibfnamefont {J.-G.}\
  \bibnamefont {Guo}}, \ and\ \bibinfo {author} {\bibfnamefont
  {X.}~\bibnamefont {Chen}},\ }\href {\doibase 10.1088/0256-307x/38/5/057402}
  {\bibfield  {journal} {\bibinfo  {journal} {Chinese Physics Letters}\
  }\textbf {\bibinfo {volume} {38}},\ \bibinfo {pages} {057402} (\bibinfo
  {year} {2021}{\natexlab{c}})}\BibitemShut {NoStop}%
\bibitem [{\citenamefont {Tan}\ \emph {et~al.}(2021)\citenamefont {Tan},
  \citenamefont {Liu}, \citenamefont {Wang},\ and\ \citenamefont
  {Yan}}]{Tan2021}%
  \BibitemOpen
  \bibfield  {author} {\bibinfo {author} {\bibfnamefont {H.}~\bibnamefont
  {Tan}}, \bibinfo {author} {\bibfnamefont {Y.}~\bibnamefont {Liu}}, \bibinfo
  {author} {\bibfnamefont {Z.}~\bibnamefont {Wang}}, \ and\ \bibinfo {author}
  {\bibfnamefont {B.}~\bibnamefont {Yan}},\ }\href {\doibase
  10.1103/physrevlett.127.046401} {\bibfield  {journal} {\bibinfo  {journal}
  {Physical Review Letters}\ }\textbf {\bibinfo {volume} {127}},\ \bibinfo
  {pages} {046401} (\bibinfo {year} {2021})}\BibitemShut {NoStop}%
\bibitem [{\citenamefont {Ortiz}\ \emph {et~al.}(2020)\citenamefont {Ortiz},
  \citenamefont {Teicher}, \citenamefont {Hu}, \citenamefont {Zuo},
  \citenamefont {Sarte}, \citenamefont {Schueller}, \citenamefont {Abeykoon},
  \citenamefont {Krogstad}, \citenamefont {Rosenkranz}, \citenamefont {Osborn},
  \citenamefont {Seshadri}, \citenamefont {Balents}, \citenamefont {He},\ and\
  \citenamefont {Wilson}}]{Ortiz2020}%
  \BibitemOpen
  \bibfield  {author} {\bibinfo {author} {\bibfnamefont {B.~R.}\ \bibnamefont
  {Ortiz}}, \bibinfo {author} {\bibfnamefont {S.~M.}\ \bibnamefont {Teicher}},
  \bibinfo {author} {\bibfnamefont {Y.}~\bibnamefont {Hu}}, \bibinfo {author}
  {\bibfnamefont {J.~L.}\ \bibnamefont {Zuo}}, \bibinfo {author} {\bibfnamefont
  {P.~M.}\ \bibnamefont {Sarte}}, \bibinfo {author} {\bibfnamefont {E.~C.}\
  \bibnamefont {Schueller}}, \bibinfo {author} {\bibfnamefont {A.~M.}\
  \bibnamefont {Abeykoon}}, \bibinfo {author} {\bibfnamefont {M.~J.}\
  \bibnamefont {Krogstad}}, \bibinfo {author} {\bibfnamefont {S.}~\bibnamefont
  {Rosenkranz}}, \bibinfo {author} {\bibfnamefont {R.}~\bibnamefont {Osborn}},
  \bibinfo {author} {\bibfnamefont {R.}~\bibnamefont {Seshadri}}, \bibinfo
  {author} {\bibfnamefont {L.}~\bibnamefont {Balents}}, \bibinfo {author}
  {\bibfnamefont {J.}~\bibnamefont {He}}, \ and\ \bibinfo {author}
  {\bibfnamefont {S.~D.}\ \bibnamefont {Wilson}},\ }\href {\doibase
  10.1103/physrevlett.125.247002} {\bibfield  {journal} {\bibinfo  {journal}
  {Physical Review Letters}\ }\textbf {\bibinfo {volume} {125}},\ \bibinfo
  {pages} {247002} (\bibinfo {year} {2020})}\BibitemShut {NoStop}%
\bibitem [{\citenamefont {Mielke}\ \emph {et~al.}(2022)\citenamefont {Mielke},
  \citenamefont {Das}, \citenamefont {Yin}, \citenamefont {Liu}, \citenamefont
  {Gupta}, \citenamefont {Jiang}, \citenamefont {Medarde}, \citenamefont {Wu},
  \citenamefont {Lei}, \citenamefont {Chang}, \citenamefont {Dai},
  \citenamefont {Si}, \citenamefont {Miao}, \citenamefont {Thomale},
  \citenamefont {Neupert}, \citenamefont {Shi}, \citenamefont {Khasanov},
  \citenamefont {Hasan}, \citenamefont {Luetkens},\ and\ \citenamefont
  {Guguchia}}]{Mielke2022}%
  \BibitemOpen
  \bibfield  {author} {\bibinfo {author} {\bibfnamefont {C.}~\bibnamefont
  {Mielke}}, \bibinfo {author} {\bibfnamefont {D.}~\bibnamefont {Das}},
  \bibinfo {author} {\bibfnamefont {J.-X.}\ \bibnamefont {Yin}}, \bibinfo
  {author} {\bibfnamefont {H.}~\bibnamefont {Liu}}, \bibinfo {author}
  {\bibfnamefont {R.}~\bibnamefont {Gupta}}, \bibinfo {author} {\bibfnamefont
  {Y.-X.}\ \bibnamefont {Jiang}}, \bibinfo {author} {\bibfnamefont
  {M.}~\bibnamefont {Medarde}}, \bibinfo {author} {\bibfnamefont
  {X.}~\bibnamefont {Wu}}, \bibinfo {author} {\bibfnamefont {H.~C.}\
  \bibnamefont {Lei}}, \bibinfo {author} {\bibfnamefont {J.}~\bibnamefont
  {Chang}}, \bibinfo {author} {\bibfnamefont {P.}~\bibnamefont {Dai}}, \bibinfo
  {author} {\bibfnamefont {Q.}~\bibnamefont {Si}}, \bibinfo {author}
  {\bibfnamefont {H.}~\bibnamefont {Miao}}, \bibinfo {author} {\bibfnamefont
  {R.}~\bibnamefont {Thomale}}, \bibinfo {author} {\bibfnamefont
  {T.}~\bibnamefont {Neupert}}, \bibinfo {author} {\bibfnamefont
  {Y.}~\bibnamefont {Shi}}, \bibinfo {author} {\bibfnamefont {R.}~\bibnamefont
  {Khasanov}}, \bibinfo {author} {\bibfnamefont {M.~Z.}\ \bibnamefont {Hasan}},
  \bibinfo {author} {\bibfnamefont {H.}~\bibnamefont {Luetkens}}, \ and\
  \bibinfo {author} {\bibfnamefont {Z.}~\bibnamefont {Guguchia}},\ }\href
  {\doibase 10.1038/s41586-021-04327-z} {\bibfield  {journal} {\bibinfo
  {journal} {Nature}\ }\textbf {\bibinfo {volume} {602}},\ \bibinfo {pages}
  {245} (\bibinfo {year} {2022})}\BibitemShut {NoStop}%
\bibitem [{\citenamefont {Jiang}\ \emph {et~al.}(2021)\citenamefont {Jiang},
  \citenamefont {Yin}, \citenamefont {Denner}, \citenamefont {Shumiya},
  \citenamefont {Ortiz}, \citenamefont {Xu}, \citenamefont {Guguchia},
  \citenamefont {He}, \citenamefont {Hossain}, \citenamefont {Liu},
  \citenamefont {Ruff}, \citenamefont {Kautzsch}, \citenamefont {Zhang},
  \citenamefont {Chang}, \citenamefont {Belopolski}, \citenamefont {Zhang},
  \citenamefont {Cochran}, \citenamefont {Multer}, \citenamefont {Litskevich},
  \citenamefont {Cheng}, \citenamefont {Yang}, \citenamefont {Wang},
  \citenamefont {Thomale}, \citenamefont {Neupert}, \citenamefont {Wilson},\
  and\ \citenamefont {Hasan}}]{Jiang2021}%
  \BibitemOpen
  \bibfield  {author} {\bibinfo {author} {\bibfnamefont {Y.-X.}\ \bibnamefont
  {Jiang}}, \bibinfo {author} {\bibfnamefont {J.-X.}\ \bibnamefont {Yin}},
  \bibinfo {author} {\bibfnamefont {M.~M.}\ \bibnamefont {Denner}}, \bibinfo
  {author} {\bibfnamefont {N.}~\bibnamefont {Shumiya}}, \bibinfo {author}
  {\bibfnamefont {B.~R.}\ \bibnamefont {Ortiz}}, \bibinfo {author}
  {\bibfnamefont {G.}~\bibnamefont {Xu}}, \bibinfo {author} {\bibfnamefont
  {Z.}~\bibnamefont {Guguchia}}, \bibinfo {author} {\bibfnamefont
  {J.}~\bibnamefont {He}}, \bibinfo {author} {\bibfnamefont {M.~S.}\
  \bibnamefont {Hossain}}, \bibinfo {author} {\bibfnamefont {X.}~\bibnamefont
  {Liu}}, \bibinfo {author} {\bibfnamefont {J.}~\bibnamefont {Ruff}}, \bibinfo
  {author} {\bibfnamefont {L.}~\bibnamefont {Kautzsch}}, \bibinfo {author}
  {\bibfnamefont {S.~S.}\ \bibnamefont {Zhang}}, \bibinfo {author}
  {\bibfnamefont {G.}~\bibnamefont {Chang}}, \bibinfo {author} {\bibfnamefont
  {I.}~\bibnamefont {Belopolski}}, \bibinfo {author} {\bibfnamefont
  {Q.}~\bibnamefont {Zhang}}, \bibinfo {author} {\bibfnamefont {T.~A.}\
  \bibnamefont {Cochran}}, \bibinfo {author} {\bibfnamefont {D.}~\bibnamefont
  {Multer}}, \bibinfo {author} {\bibfnamefont {M.}~\bibnamefont {Litskevich}},
  \bibinfo {author} {\bibfnamefont {Z.-J.}\ \bibnamefont {Cheng}}, \bibinfo
  {author} {\bibfnamefont {X.~P.}\ \bibnamefont {Yang}}, \bibinfo {author}
  {\bibfnamefont {Z.}~\bibnamefont {Wang}}, \bibinfo {author} {\bibfnamefont
  {R.}~\bibnamefont {Thomale}}, \bibinfo {author} {\bibfnamefont
  {T.}~\bibnamefont {Neupert}}, \bibinfo {author} {\bibfnamefont {S.~D.}\
  \bibnamefont {Wilson}}, \ and\ \bibinfo {author} {\bibfnamefont {M.~Z.}\
  \bibnamefont {Hasan}},\ }\href {\doibase 10.1038/s41563-021-01034-y}
  {\bibfield  {journal} {\bibinfo  {journal} {Nature Materials}\ }\textbf
  {\bibinfo {volume} {20}},\ \bibinfo {pages} {1353} (\bibinfo {year}
  {2021})}\BibitemShut {NoStop}%
\bibitem [{\citenamefont {Yang}\ \emph {et~al.}(2020)\citenamefont {Yang},
  \citenamefont {Wang}, \citenamefont {Ortiz}, \citenamefont {Liu},
  \citenamefont {Gayles}, \citenamefont {Derunova}, \citenamefont
  {Gonzalez-Hernandez}, \citenamefont {Šmejkal}, \citenamefont {Chen},
  \citenamefont {Parkin}, \citenamefont {Wilson}, \citenamefont {Toberer},
  \citenamefont {McQueen},\ and\ \citenamefont {Ali}}]{Yang2020}%
  \BibitemOpen
  \bibfield  {author} {\bibinfo {author} {\bibfnamefont {S.-Y.}\ \bibnamefont
  {Yang}}, \bibinfo {author} {\bibfnamefont {Y.}~\bibnamefont {Wang}}, \bibinfo
  {author} {\bibfnamefont {B.~R.}\ \bibnamefont {Ortiz}}, \bibinfo {author}
  {\bibfnamefont {D.}~\bibnamefont {Liu}}, \bibinfo {author} {\bibfnamefont
  {J.}~\bibnamefont {Gayles}}, \bibinfo {author} {\bibfnamefont
  {E.}~\bibnamefont {Derunova}}, \bibinfo {author} {\bibfnamefont
  {R.}~\bibnamefont {Gonzalez-Hernandez}}, \bibinfo {author} {\bibfnamefont
  {L.}~\bibnamefont {Šmejkal}}, \bibinfo {author} {\bibfnamefont
  {Y.}~\bibnamefont {Chen}}, \bibinfo {author} {\bibfnamefont {S.~S.~P.}\
  \bibnamefont {Parkin}}, \bibinfo {author} {\bibfnamefont {S.~D.}\
  \bibnamefont {Wilson}}, \bibinfo {author} {\bibfnamefont {E.~S.}\
  \bibnamefont {Toberer}}, \bibinfo {author} {\bibfnamefont {T.}~\bibnamefont
  {McQueen}}, \ and\ \bibinfo {author} {\bibfnamefont {M.~N.}\ \bibnamefont
  {Ali}},\ }\href {\doibase 10.1126/sciadv.abb6003} {\bibfield  {journal}
  {\bibinfo  {journal} {Science Advances}\ }\textbf {\bibinfo {volume} {6}},\
  \bibinfo {pages} {eabb6003} (\bibinfo {year} {2020})}\BibitemShut {NoStop}%
\bibitem [{\citenamefont {Hu}\ \emph {et~al.}(2022{\natexlab{a}})\citenamefont
  {Hu}, \citenamefont {Teicher}, \citenamefont {Ortiz}, \citenamefont {Luo},
  \citenamefont {Peng}, \citenamefont {Huai}, \citenamefont {Ma}, \citenamefont
  {Plumb}, \citenamefont {Wilson}, \citenamefont {He},\ and\ \citenamefont
  {Shi}}]{Hu2022}%
  \BibitemOpen
  \bibfield  {author} {\bibinfo {author} {\bibfnamefont {Y.}~\bibnamefont
  {Hu}}, \bibinfo {author} {\bibfnamefont {S.~M.}\ \bibnamefont {Teicher}},
  \bibinfo {author} {\bibfnamefont {B.~R.}\ \bibnamefont {Ortiz}}, \bibinfo
  {author} {\bibfnamefont {Y.}~\bibnamefont {Luo}}, \bibinfo {author}
  {\bibfnamefont {S.}~\bibnamefont {Peng}}, \bibinfo {author} {\bibfnamefont
  {L.}~\bibnamefont {Huai}}, \bibinfo {author} {\bibfnamefont {J.}~\bibnamefont
  {Ma}}, \bibinfo {author} {\bibfnamefont {N.~C.}\ \bibnamefont {Plumb}},
  \bibinfo {author} {\bibfnamefont {S.~D.}\ \bibnamefont {Wilson}}, \bibinfo
  {author} {\bibfnamefont {J.}~\bibnamefont {He}}, \ and\ \bibinfo {author}
  {\bibfnamefont {M.}~\bibnamefont {Shi}},\ }\href {\doibase
  10.1016/j.scib.2021.11.026} {\bibfield  {journal} {\bibinfo  {journal}
  {Science Bulletin}\ }\textbf {\bibinfo {volume} {67}},\ \bibinfo {pages}
  {495} (\bibinfo {year} {2022}{\natexlab{a}})}\BibitemShut {NoStop}%
\bibitem [{\citenamefont {Lin}\ and\ \citenamefont
  {Nandkishore}(2021)}]{Lin2021}%
  \BibitemOpen
  \bibfield  {author} {\bibinfo {author} {\bibfnamefont {Y.-P.}\ \bibnamefont
  {Lin}}\ and\ \bibinfo {author} {\bibfnamefont {R.~M.}\ \bibnamefont
  {Nandkishore}},\ }\href {\doibase 10.1103/physrevb.104.045122} {\bibfield
  {journal} {\bibinfo  {journal} {Physical Review B}\ }\textbf {\bibinfo
  {volume} {104}},\ \bibinfo {pages} {045122} (\bibinfo {year}
  {2021})}\BibitemShut {NoStop}%
\bibitem [{\citenamefont {Park}\ \emph {et~al.}(2021)\citenamefont {Park},
  \citenamefont {Ye},\ and\ \citenamefont {Balents}}]{Park2021}%
  \BibitemOpen
  \bibfield  {author} {\bibinfo {author} {\bibfnamefont {T.}~\bibnamefont
  {Park}}, \bibinfo {author} {\bibfnamefont {M.}~\bibnamefont {Ye}}, \ and\
  \bibinfo {author} {\bibfnamefont {L.}~\bibnamefont {Balents}},\ }\href
  {\doibase 10.1103/physrevb.104.035142} {\bibfield  {journal} {\bibinfo
  {journal} {Physical Review B}\ }\textbf {\bibinfo {volume} {104}},\ \bibinfo
  {pages} {035142} (\bibinfo {year} {2021})}\BibitemShut {NoStop}%
\bibitem [{\citenamefont {Feng}\ \emph {et~al.}(2021)\citenamefont {Feng},
  \citenamefont {Jiang}, \citenamefont {Wang},\ and\ \citenamefont
  {Hu}}]{Feng2021}%
  \BibitemOpen
  \bibfield  {author} {\bibinfo {author} {\bibfnamefont {X.}~\bibnamefont
  {Feng}}, \bibinfo {author} {\bibfnamefont {K.}~\bibnamefont {Jiang}},
  \bibinfo {author} {\bibfnamefont {Z.}~\bibnamefont {Wang}}, \ and\ \bibinfo
  {author} {\bibfnamefont {J.}~\bibnamefont {Hu}},\ }\href {\doibase
  10.1016/j.scib.2021.04.043} {\bibfield  {journal} {\bibinfo  {journal}
  {Science Bulletin}\ }\textbf {\bibinfo {volume} {66}},\ \bibinfo {pages}
  {1384} (\bibinfo {year} {2021})}\BibitemShut {NoStop}%
\bibitem [{\citenamefont {Wu}\ \emph {et~al.}(2021)\citenamefont {Wu},
  \citenamefont {Schwemmer}, \citenamefont {Müller}, \citenamefont
  {Consiglio}, \citenamefont {Sangiovanni}, \citenamefont {Di~Sante},
  \citenamefont {Iqbal}, \citenamefont {Hanke}, \citenamefont {Schnyder},
  \citenamefont {Denner}, \citenamefont {Fischer}, \citenamefont {Neupert},\
  and\ \citenamefont {Thomale}}]{Wu2021}%
  \BibitemOpen
  \bibfield  {author} {\bibinfo {author} {\bibfnamefont {X.}~\bibnamefont
  {Wu}}, \bibinfo {author} {\bibfnamefont {T.}~\bibnamefont {Schwemmer}},
  \bibinfo {author} {\bibfnamefont {T.}~\bibnamefont {Müller}}, \bibinfo
  {author} {\bibfnamefont {A.}~\bibnamefont {Consiglio}}, \bibinfo {author}
  {\bibfnamefont {G.}~\bibnamefont {Sangiovanni}}, \bibinfo {author}
  {\bibfnamefont {D.}~\bibnamefont {Di~Sante}}, \bibinfo {author}
  {\bibfnamefont {Y.}~\bibnamefont {Iqbal}}, \bibinfo {author} {\bibfnamefont
  {W.}~\bibnamefont {Hanke}}, \bibinfo {author} {\bibfnamefont {A.~P.}\
  \bibnamefont {Schnyder}}, \bibinfo {author} {\bibfnamefont {M.~M.}\
  \bibnamefont {Denner}}, \bibinfo {author} {\bibfnamefont {M.~H.}\
  \bibnamefont {Fischer}}, \bibinfo {author} {\bibfnamefont {T.}~\bibnamefont
  {Neupert}}, \ and\ \bibinfo {author} {\bibfnamefont {R.}~\bibnamefont
  {Thomale}},\ }\href {\doibase 10.1103/physrevlett.127.177001} {\bibfield
  {journal} {\bibinfo  {journal} {Physical Review Letters}\ }\textbf {\bibinfo
  {volume} {127}},\ \bibinfo {pages} {177001} (\bibinfo {year}
  {2021})}\BibitemShut {NoStop}%
\bibitem [{\citenamefont {Denner}\ \emph {et~al.}(2021)\citenamefont {Denner},
  \citenamefont {Thomale},\ and\ \citenamefont {Neupert}}]{Denner2021}%
  \BibitemOpen
  \bibfield  {author} {\bibinfo {author} {\bibfnamefont {M.~M.}\ \bibnamefont
  {Denner}}, \bibinfo {author} {\bibfnamefont {R.}~\bibnamefont {Thomale}}, \
  and\ \bibinfo {author} {\bibfnamefont {T.}~\bibnamefont {Neupert}},\ }\href
  {\doibase 10.1103/physrevlett.127.217601} {\bibfield  {journal} {\bibinfo
  {journal} {Physical Review Letters}\ }\textbf {\bibinfo {volume} {127}},\
  \bibinfo {pages} {217601} (\bibinfo {year} {2021})}\BibitemShut {NoStop}%
\bibitem [{\citenamefont {Ortiz}\ \emph
  {et~al.}(2021{\natexlab{a}})\citenamefont {Ortiz}, \citenamefont {Sarte},
  \citenamefont {Kenney}, \citenamefont {Graf}, \citenamefont {Teicher},
  \citenamefont {Seshadri},\ and\ \citenamefont {Wilson}}]{Ortiz2021}%
  \BibitemOpen
  \bibfield  {author} {\bibinfo {author} {\bibfnamefont {B.~R.}\ \bibnamefont
  {Ortiz}}, \bibinfo {author} {\bibfnamefont {P.~M.}\ \bibnamefont {Sarte}},
  \bibinfo {author} {\bibfnamefont {E.~M.}\ \bibnamefont {Kenney}}, \bibinfo
  {author} {\bibfnamefont {M.~J.}\ \bibnamefont {Graf}}, \bibinfo {author}
  {\bibfnamefont {S.~M.~L.}\ \bibnamefont {Teicher}}, \bibinfo {author}
  {\bibfnamefont {R.}~\bibnamefont {Seshadri}}, \ and\ \bibinfo {author}
  {\bibfnamefont {S.~D.}\ \bibnamefont {Wilson}},\ }\href {\doibase
  10.1103/physrevmaterials.5.034801} {\bibfield  {journal} {\bibinfo  {journal}
  {Physical Review Materials}\ }\textbf {\bibinfo {volume} {5}},\ \bibinfo
  {pages} {034801} (\bibinfo {year} {2021}{\natexlab{a}})}\BibitemShut
  {NoStop}%
\bibitem [{\citenamefont {Wei}\ \emph {et~al.}(2024)\citenamefont {Wei},
  \citenamefont {Tian}, \citenamefont {Cui}, \citenamefont {Zhai},
  \citenamefont {Li}, \citenamefont {Liu}, \citenamefont {Song}, \citenamefont
  {Feng}, \citenamefont {Huang}, \citenamefont {Wang}, \citenamefont {Liu},
  \citenamefont {Xiong}, \citenamefont {Yao}, \citenamefont {Xie},\ and\
  \citenamefont {Chen}}]{Wei2024}%
  \BibitemOpen
  \bibfield  {author} {\bibinfo {author} {\bibfnamefont {X.}~\bibnamefont
  {Wei}}, \bibinfo {author} {\bibfnamefont {C.}~\bibnamefont {Tian}}, \bibinfo
  {author} {\bibfnamefont {H.}~\bibnamefont {Cui}}, \bibinfo {author}
  {\bibfnamefont {Y.}~\bibnamefont {Zhai}}, \bibinfo {author} {\bibfnamefont
  {Y.}~\bibnamefont {Li}}, \bibinfo {author} {\bibfnamefont {S.}~\bibnamefont
  {Liu}}, \bibinfo {author} {\bibfnamefont {Y.}~\bibnamefont {Song}}, \bibinfo
  {author} {\bibfnamefont {Y.}~\bibnamefont {Feng}}, \bibinfo {author}
  {\bibfnamefont {M.}~\bibnamefont {Huang}}, \bibinfo {author} {\bibfnamefont
  {Z.}~\bibnamefont {Wang}}, \bibinfo {author} {\bibfnamefont {Y.}~\bibnamefont
  {Liu}}, \bibinfo {author} {\bibfnamefont {Q.}~\bibnamefont {Xiong}}, \bibinfo
  {author} {\bibfnamefont {Y.}~\bibnamefont {Yao}}, \bibinfo {author}
  {\bibfnamefont {X.~C.}\ \bibnamefont {Xie}}, \ and\ \bibinfo {author}
  {\bibfnamefont {J.-H.}\ \bibnamefont {Chen}},\ }\href {\doibase
  10.1038/s41467-024-49248-3} {\bibfield  {journal} {\bibinfo  {journal}
  {Nature Communications}\ }\textbf {\bibinfo {volume} {15}},\ \bibinfo {pages}
  {5038} (\bibinfo {year} {2024})}\BibitemShut {NoStop}%
\bibitem [{\citenamefont {Zhou}\ \emph {et~al.}(2022)\citenamefont {Zhou},
  \citenamefont {Liu}, \citenamefont {Wu}, \citenamefont {Jiang}, \citenamefont
  {Shi}, \citenamefont {Li}, \citenamefont {Sui}, \citenamefont {Hu},\ and\
  \citenamefont {Luo}}]{Zhou2022}%
  \BibitemOpen
  \bibfield  {author} {\bibinfo {author} {\bibfnamefont {X.}~\bibnamefont
  {Zhou}}, \bibinfo {author} {\bibfnamefont {H.}~\bibnamefont {Liu}}, \bibinfo
  {author} {\bibfnamefont {W.}~\bibnamefont {Wu}}, \bibinfo {author}
  {\bibfnamefont {K.}~\bibnamefont {Jiang}}, \bibinfo {author} {\bibfnamefont
  {Y.}~\bibnamefont {Shi}}, \bibinfo {author} {\bibfnamefont {Z.}~\bibnamefont
  {Li}}, \bibinfo {author} {\bibfnamefont {Y.}~\bibnamefont {Sui}}, \bibinfo
  {author} {\bibfnamefont {J.}~\bibnamefont {Hu}}, \ and\ \bibinfo {author}
  {\bibfnamefont {J.}~\bibnamefont {Luo}},\ }\href {\doibase
  10.1103/physrevb.105.205104} {\bibfield  {journal} {\bibinfo  {journal}
  {Physical Review B}\ }\textbf {\bibinfo {volume} {105}},\ \bibinfo {pages}
  {205104} (\bibinfo {year} {2022})}\BibitemShut {NoStop}%
\bibitem [{\citenamefont {Yu}\ \emph {et~al.}(2021)\citenamefont {Yu},
  \citenamefont {Wu}, \citenamefont {Wang}, \citenamefont {Lei}, \citenamefont
  {Zhuo}, \citenamefont {Ying},\ and\ \citenamefont {Chen}}]{Yu2021}%
  \BibitemOpen
  \bibfield  {author} {\bibinfo {author} {\bibfnamefont {F.~H.}\ \bibnamefont
  {Yu}}, \bibinfo {author} {\bibfnamefont {T.}~\bibnamefont {Wu}}, \bibinfo
  {author} {\bibfnamefont {Z.~Y.}\ \bibnamefont {Wang}}, \bibinfo {author}
  {\bibfnamefont {B.}~\bibnamefont {Lei}}, \bibinfo {author} {\bibfnamefont
  {W.~Z.}\ \bibnamefont {Zhuo}}, \bibinfo {author} {\bibfnamefont {J.~J.}\
  \bibnamefont {Ying}}, \ and\ \bibinfo {author} {\bibfnamefont {X.~H.}\
  \bibnamefont {Chen}},\ }\href {\doibase 10.1103/physrevb.104.l041103}
  {\bibfield  {journal} {\bibinfo  {journal} {Physical Review B}\ }\textbf
  {\bibinfo {volume} {104}},\ \bibinfo {pages} {L041103} (\bibinfo {year}
  {2021})}\BibitemShut {NoStop}%
\bibitem [{\citenamefont {Khasanov}\ \emph {et~al.}(2022)\citenamefont
  {Khasanov}, \citenamefont {Das}, \citenamefont {Gupta}, \citenamefont
  {Mielke}, \citenamefont {Elender}, \citenamefont {Yin}, \citenamefont {Tu},
  \citenamefont {Gong}, \citenamefont {Lei}, \citenamefont {Ritz},
  \citenamefont {Fernandes}, \citenamefont {Birol}, \citenamefont {Guguchia},\
  and\ \citenamefont {Luetkens}}]{Khasanov2022}%
  \BibitemOpen
  \bibfield  {author} {\bibinfo {author} {\bibfnamefont {R.}~\bibnamefont
  {Khasanov}}, \bibinfo {author} {\bibfnamefont {D.}~\bibnamefont {Das}},
  \bibinfo {author} {\bibfnamefont {R.}~\bibnamefont {Gupta}}, \bibinfo
  {author} {\bibfnamefont {C.}~\bibnamefont {Mielke}}, \bibinfo {author}
  {\bibfnamefont {M.}~\bibnamefont {Elender}}, \bibinfo {author} {\bibfnamefont
  {Q.}~\bibnamefont {Yin}}, \bibinfo {author} {\bibfnamefont {Z.}~\bibnamefont
  {Tu}}, \bibinfo {author} {\bibfnamefont {C.}~\bibnamefont {Gong}}, \bibinfo
  {author} {\bibfnamefont {H.}~\bibnamefont {Lei}}, \bibinfo {author}
  {\bibfnamefont {E.~T.}\ \bibnamefont {Ritz}}, \bibinfo {author}
  {\bibfnamefont {R.~M.}\ \bibnamefont {Fernandes}}, \bibinfo {author}
  {\bibfnamefont {T.}~\bibnamefont {Birol}}, \bibinfo {author} {\bibfnamefont
  {Z.}~\bibnamefont {Guguchia}}, \ and\ \bibinfo {author} {\bibfnamefont
  {H.}~\bibnamefont {Luetkens}},\ }\href {\doibase
  10.1103/physrevresearch.4.023244} {\bibfield  {journal} {\bibinfo  {journal}
  {Physical Review Research}\ }\textbf {\bibinfo {volume} {4}},\ \bibinfo
  {pages} {023244} (\bibinfo {year} {2022})}\BibitemShut {NoStop}%
\bibitem [{\citenamefont {Hu}\ \emph {et~al.}(2022{\natexlab{b}})\citenamefont
  {Hu}, \citenamefont {Yamane}, \citenamefont {Mattoni}, \citenamefont {Yada},
  \citenamefont {Obata}, \citenamefont {Li}, \citenamefont {Yao}, \citenamefont
  {Wang}, \citenamefont {Wang}, \citenamefont {Farhang}, \citenamefont {Xia},
  \citenamefont {Maeno},\ and\ \citenamefont {Yonezawa}}]{Hu2022a}%
  \BibitemOpen
  \bibfield  {author} {\bibinfo {author} {\bibfnamefont {Y.}~\bibnamefont
  {Hu}}, \bibinfo {author} {\bibfnamefont {S.}~\bibnamefont {Yamane}}, \bibinfo
  {author} {\bibfnamefont {G.}~\bibnamefont {Mattoni}}, \bibinfo {author}
  {\bibfnamefont {K.}~\bibnamefont {Yada}}, \bibinfo {author} {\bibfnamefont
  {K.}~\bibnamefont {Obata}}, \bibinfo {author} {\bibfnamefont
  {Y.}~\bibnamefont {Li}}, \bibinfo {author} {\bibfnamefont {Y.}~\bibnamefont
  {Yao}}, \bibinfo {author} {\bibfnamefont {Z.}~\bibnamefont {Wang}}, \bibinfo
  {author} {\bibfnamefont {J.}~\bibnamefont {Wang}}, \bibinfo {author}
  {\bibfnamefont {C.}~\bibnamefont {Farhang}}, \bibinfo {author} {\bibfnamefont
  {J.}~\bibnamefont {Xia}}, \bibinfo {author} {\bibfnamefont {Y.}~\bibnamefont
  {Maeno}}, \ and\ \bibinfo {author} {\bibfnamefont {S.}~\bibnamefont
  {Yonezawa}},\ }\href {\doibase 10.48550/ARXIV.2208.08036} {\enquote {\bibinfo
  {title} {Time-reversal symmetry breaking in charge density wave of
  csv$_3$sb$_5$ detected by polar kerr effect},}\ }\bibinfo {howpublished}
  {arXiv:2208.08036v2} (\bibinfo {year} {2022}{\natexlab{b}})\BibitemShut
  {NoStop}%
\bibitem [{\citenamefont {Wu}\ \emph {et~al.}(2022)\citenamefont {Wu},
  \citenamefont {Wang}, \citenamefont {Liu}, \citenamefont {Li}, \citenamefont
  {Xu}, \citenamefont {Yin}, \citenamefont {Gong}, \citenamefont {Tu},
  \citenamefont {Lei}, \citenamefont {Dong},\ and\ \citenamefont
  {Wang}}]{Wu2022}%
  \BibitemOpen
  \bibfield  {author} {\bibinfo {author} {\bibfnamefont {Q.}~\bibnamefont
  {Wu}}, \bibinfo {author} {\bibfnamefont {Z.~X.}\ \bibnamefont {Wang}},
  \bibinfo {author} {\bibfnamefont {Q.~M.}\ \bibnamefont {Liu}}, \bibinfo
  {author} {\bibfnamefont {R.~S.}\ \bibnamefont {Li}}, \bibinfo {author}
  {\bibfnamefont {S.~X.}\ \bibnamefont {Xu}}, \bibinfo {author} {\bibfnamefont
  {Q.~W.}\ \bibnamefont {Yin}}, \bibinfo {author} {\bibfnamefont {C.~S.}\
  \bibnamefont {Gong}}, \bibinfo {author} {\bibfnamefont {Z.~J.}\ \bibnamefont
  {Tu}}, \bibinfo {author} {\bibfnamefont {H.~C.}\ \bibnamefont {Lei}},
  \bibinfo {author} {\bibfnamefont {T.}~\bibnamefont {Dong}}, \ and\ \bibinfo
  {author} {\bibfnamefont {N.~L.}\ \bibnamefont {Wang}},\ }\href {\doibase
  10.1103/physrevb.106.205109} {\bibfield  {journal} {\bibinfo  {journal}
  {Physical Review B}\ }\textbf {\bibinfo {volume} {106}},\ \bibinfo {pages}
  {205109} (\bibinfo {year} {2022})}\BibitemShut {NoStop}%
\bibitem [{\citenamefont {Farhang}\ \emph {et~al.}(2023)\citenamefont
  {Farhang}, \citenamefont {Wang}, \citenamefont {Ortiz}, \citenamefont
  {Wilson},\ and\ \citenamefont {Xia}}]{Farhang2023}%
  \BibitemOpen
  \bibfield  {author} {\bibinfo {author} {\bibfnamefont {C.}~\bibnamefont
  {Farhang}}, \bibinfo {author} {\bibfnamefont {J.}~\bibnamefont {Wang}},
  \bibinfo {author} {\bibfnamefont {B.~R.}\ \bibnamefont {Ortiz}}, \bibinfo
  {author} {\bibfnamefont {S.~D.}\ \bibnamefont {Wilson}}, \ and\ \bibinfo
  {author} {\bibfnamefont {J.}~\bibnamefont {Xia}},\ }\href {\doibase
  10.1038/s41467-023-41080-5} {\bibfield  {journal} {\bibinfo  {journal}
  {Nature Communications}\ }\textbf {\bibinfo {volume} {14}},\ \bibinfo {pages}
  {5326} (\bibinfo {year} {2023})}\BibitemShut {NoStop}%
\bibitem [{\citenamefont {Wang}\ \emph {et~al.}(2024)\citenamefont {Wang},
  \citenamefont {Farhang}, \citenamefont {Ortiz}, \citenamefont {Wilson},\ and\
  \citenamefont {Xia}}]{Wang2024}%
  \BibitemOpen
  \bibfield  {author} {\bibinfo {author} {\bibfnamefont {J.}~\bibnamefont
  {Wang}}, \bibinfo {author} {\bibfnamefont {C.}~\bibnamefont {Farhang}},
  \bibinfo {author} {\bibfnamefont {B.~R.}\ \bibnamefont {Ortiz}}, \bibinfo
  {author} {\bibfnamefont {S.~D.}\ \bibnamefont {Wilson}}, \ and\ \bibinfo
  {author} {\bibfnamefont {J.}~\bibnamefont {Xia}},\ }\href {\doibase
  10.1103/physrevmaterials.8.014202} {\bibfield  {journal} {\bibinfo  {journal}
  {Physical Review Materials}\ }\textbf {\bibinfo {volume} {8}},\ \bibinfo
  {pages} {014202} (\bibinfo {year} {2024})}\BibitemShut {NoStop}%
\bibitem [{\citenamefont {Saykin}\ \emph {et~al.}(2023)\citenamefont {Saykin},
  \citenamefont {Farhang}, \citenamefont {Kountz}, \citenamefont {Chen},
  \citenamefont {Ortiz}, \citenamefont {Shekhar}, \citenamefont {Felser},
  \citenamefont {Wilson}, \citenamefont {Thomale}, \citenamefont {Xia},\ and\
  \citenamefont {Kapitulnik}}]{Saykin2023}%
  \BibitemOpen
  \bibfield  {author} {\bibinfo {author} {\bibfnamefont {D.~R.}\ \bibnamefont
  {Saykin}}, \bibinfo {author} {\bibfnamefont {C.}~\bibnamefont {Farhang}},
  \bibinfo {author} {\bibfnamefont {E.~D.}\ \bibnamefont {Kountz}}, \bibinfo
  {author} {\bibfnamefont {D.}~\bibnamefont {Chen}}, \bibinfo {author}
  {\bibfnamefont {B.~R.}\ \bibnamefont {Ortiz}}, \bibinfo {author}
  {\bibfnamefont {C.}~\bibnamefont {Shekhar}}, \bibinfo {author} {\bibfnamefont
  {C.}~\bibnamefont {Felser}}, \bibinfo {author} {\bibfnamefont {S.~D.}\
  \bibnamefont {Wilson}}, \bibinfo {author} {\bibfnamefont {R.}~\bibnamefont
  {Thomale}}, \bibinfo {author} {\bibfnamefont {J.}~\bibnamefont {Xia}}, \ and\
  \bibinfo {author} {\bibfnamefont {A.}~\bibnamefont {Kapitulnik}},\ }\href
  {\doibase 10.1103/physrevlett.131.016901} {\bibfield  {journal} {\bibinfo
  {journal} {Physical Review Letters}\ }\textbf {\bibinfo {volume} {131}},\
  \bibinfo {pages} {016901} (\bibinfo {year} {2023})}\BibitemShut {NoStop}%
\bibitem [{\citenamefont {Li}\ \emph {et~al.}(2022)\citenamefont {Li},
  \citenamefont {Wan}, \citenamefont {Li}, \citenamefont {Li}, \citenamefont
  {Gu}, \citenamefont {Yang}, \citenamefont {Li}, \citenamefont {Wang},
  \citenamefont {Yao},\ and\ \citenamefont {Wen}}]{Li2022}%
  \BibitemOpen
  \bibfield  {author} {\bibinfo {author} {\bibfnamefont {H.}~\bibnamefont
  {Li}}, \bibinfo {author} {\bibfnamefont {S.}~\bibnamefont {Wan}}, \bibinfo
  {author} {\bibfnamefont {H.}~\bibnamefont {Li}}, \bibinfo {author}
  {\bibfnamefont {Q.}~\bibnamefont {Li}}, \bibinfo {author} {\bibfnamefont
  {Q.}~\bibnamefont {Gu}}, \bibinfo {author} {\bibfnamefont {H.}~\bibnamefont
  {Yang}}, \bibinfo {author} {\bibfnamefont {Y.}~\bibnamefont {Li}}, \bibinfo
  {author} {\bibfnamefont {Z.}~\bibnamefont {Wang}}, \bibinfo {author}
  {\bibfnamefont {Y.}~\bibnamefont {Yao}}, \ and\ \bibinfo {author}
  {\bibfnamefont {H.-H.}\ \bibnamefont {Wen}},\ }\href {\doibase
  10.1103/physrevb.105.045102} {\bibfield  {journal} {\bibinfo  {journal}
  {Physical Review B}\ }\textbf {\bibinfo {volume} {105}},\ \bibinfo {pages}
  {045102} (\bibinfo {year} {2022})}\BibitemShut {NoStop}%
\bibitem [{\citenamefont {Stier}\ \emph {et~al.}(2024)\citenamefont {Stier},
  \citenamefont {Haghighirad}, \citenamefont {Garbarino}, \citenamefont
  {Mishra}, \citenamefont {Stilkerich}, \citenamefont {Chen}, \citenamefont
  {Shekhar}, \citenamefont {Lacmann}, \citenamefont {Felser}, \citenamefont
  {Ritschel}, \citenamefont {Geck},\ and\ \citenamefont {Tacon}}]{Stier2024}%
  \BibitemOpen
  \bibfield  {author} {\bibinfo {author} {\bibfnamefont {F.}~\bibnamefont
  {Stier}}, \bibinfo {author} {\bibfnamefont {A.~A.}\ \bibnamefont
  {Haghighirad}}, \bibinfo {author} {\bibfnamefont {G.}~\bibnamefont
  {Garbarino}}, \bibinfo {author} {\bibfnamefont {S.}~\bibnamefont {Mishra}},
  \bibinfo {author} {\bibfnamefont {N.}~\bibnamefont {Stilkerich}}, \bibinfo
  {author} {\bibfnamefont {D.}~\bibnamefont {Chen}}, \bibinfo {author}
  {\bibfnamefont {C.}~\bibnamefont {Shekhar}}, \bibinfo {author} {\bibfnamefont
  {T.}~\bibnamefont {Lacmann}}, \bibinfo {author} {\bibfnamefont
  {C.}~\bibnamefont {Felser}}, \bibinfo {author} {\bibfnamefont
  {T.}~\bibnamefont {Ritschel}}, \bibinfo {author} {\bibfnamefont
  {J.}~\bibnamefont {Geck}}, \ and\ \bibinfo {author} {\bibfnamefont {M.~L.}\
  \bibnamefont {Tacon}},\ }\href {\doibase 10.48550/ARXIV.2404.14790} {\enquote
  {\bibinfo {title} {Pressure-dependent electronic superlattice in the
  kagome-superconductor csv$\mathrm{_3}$sb$\mathrm{_5}$},}\ }\bibinfo
  {howpublished} {arXiv:2404.14790v1} (\bibinfo {year} {2024})\BibitemShut
  {NoStop}%
\bibitem [{Sup()}]{Suppl}%
  \BibitemOpen
  \href@noop {} {\enquote {\bibinfo {title} {{See Supplemental material at [URL
  will be inserted by publisher] for additional XPD data on V 2p and Sb 3d core
  levels, details on the electronic band disperssion and the discussion of CDW
  induced changes in the electronic structure near the M point of
  ${\mathrm{CsV}}_{3}{\mathrm{Sb}}_{5}$.}}}\ }\BibitemShut {NoStop}%
\bibitem [{\citenamefont {Schlueter}\ \emph {et~al.}(2019)\citenamefont
  {Schlueter}, \citenamefont {Gloskovskii}, \citenamefont {Ederer},
  \citenamefont {Schostak}, \citenamefont {Piec}, \citenamefont {Sarkar},
  \citenamefont {Matveyev}, \citenamefont {Lömker}, \citenamefont {Sing},
  \citenamefont {Claessen}, \citenamefont {Wiemann}, \citenamefont {Schneider},
  \citenamefont {Medjanik}, \citenamefont {Schönhense}, \citenamefont {Amann},
  \citenamefont {Nilsson},\ and\ \citenamefont {Drube}}]{Schlueter2019}%
  \BibitemOpen
  \bibfield  {author} {\bibinfo {author} {\bibfnamefont {C.}~\bibnamefont
  {Schlueter}}, \bibinfo {author} {\bibfnamefont {A.}~\bibnamefont
  {Gloskovskii}}, \bibinfo {author} {\bibfnamefont {K.}~\bibnamefont {Ederer}},
  \bibinfo {author} {\bibfnamefont {I.}~\bibnamefont {Schostak}}, \bibinfo
  {author} {\bibfnamefont {S.}~\bibnamefont {Piec}}, \bibinfo {author}
  {\bibfnamefont {I.}~\bibnamefont {Sarkar}}, \bibinfo {author} {\bibfnamefont
  {Y.}~\bibnamefont {Matveyev}}, \bibinfo {author} {\bibfnamefont
  {P.}~\bibnamefont {Lömker}}, \bibinfo {author} {\bibfnamefont
  {M.}~\bibnamefont {Sing}}, \bibinfo {author} {\bibfnamefont {R.}~\bibnamefont
  {Claessen}}, \bibinfo {author} {\bibfnamefont {C.}~\bibnamefont {Wiemann}},
  \bibinfo {author} {\bibfnamefont {C.~M.}\ \bibnamefont {Schneider}}, \bibinfo
  {author} {\bibfnamefont {K.}~\bibnamefont {Medjanik}}, \bibinfo {author}
  {\bibfnamefont {G.}~\bibnamefont {Schönhense}}, \bibinfo {author}
  {\bibfnamefont {P.}~\bibnamefont {Amann}}, \bibinfo {author} {\bibfnamefont
  {A.}~\bibnamefont {Nilsson}}, \ and\ \bibinfo {author} {\bibfnamefont
  {W.}~\bibnamefont {Drube}},\ }in\ \href {\doibase 10.1063/1.5084611} {\emph
  {\bibinfo {booktitle} {{AIP} Conference Proceedings}}},\ Vol.\ \bibinfo
  {volume} {2054}\ (\bibinfo  {publisher} {AIP},\ \bibinfo {year} {2019})\ p.\
  \bibinfo {pages} {040010}\BibitemShut {NoStop}%
\bibitem [{\citenamefont {Kautzsch}\ \emph {et~al.}(2023)\citenamefont
  {Kautzsch}, \citenamefont {Ortiz}, \citenamefont {Mallayya}, \citenamefont
  {Plumb}, \citenamefont {Pokharel}, \citenamefont {Ruff}, \citenamefont
  {Islam}, \citenamefont {Kim}, \citenamefont {Seshadri},\ and\ \citenamefont
  {Wilson}}]{Kautzsch2023}%
  \BibitemOpen
  \bibfield  {author} {\bibinfo {author} {\bibfnamefont {L.}~\bibnamefont
  {Kautzsch}}, \bibinfo {author} {\bibfnamefont {B.~R.}\ \bibnamefont {Ortiz}},
  \bibinfo {author} {\bibfnamefont {K.}~\bibnamefont {Mallayya}}, \bibinfo
  {author} {\bibfnamefont {J.}~\bibnamefont {Plumb}}, \bibinfo {author}
  {\bibfnamefont {G.}~\bibnamefont {Pokharel}}, \bibinfo {author}
  {\bibfnamefont {J.~P.~C.}\ \bibnamefont {Ruff}}, \bibinfo {author}
  {\bibfnamefont {Z.}~\bibnamefont {Islam}}, \bibinfo {author} {\bibfnamefont
  {E.-A.}\ \bibnamefont {Kim}}, \bibinfo {author} {\bibfnamefont
  {R.}~\bibnamefont {Seshadri}}, \ and\ \bibinfo {author} {\bibfnamefont
  {S.~D.}\ \bibnamefont {Wilson}},\ }\href {\doibase
  10.1103/physrevmaterials.7.024806} {\bibfield  {journal} {\bibinfo  {journal}
  {Physical Review Materials}\ }\textbf {\bibinfo {volume} {7}},\ \bibinfo
  {pages} {024806} (\bibinfo {year} {2023})}\BibitemShut {NoStop}%
\bibitem [{\citenamefont {Ortiz}\ \emph
  {et~al.}(2021{\natexlab{b}})\citenamefont {Ortiz}, \citenamefont {Teicher},
  \citenamefont {Kautzsch}, \citenamefont {Sarte}, \citenamefont {Ratcliff},
  \citenamefont {Harter}, \citenamefont {Ruff}, \citenamefont {Seshadri},\ and\
  \citenamefont {Wilson}}]{Ortiz2021a}%
  \BibitemOpen
  \bibfield  {author} {\bibinfo {author} {\bibfnamefont {B.~R.}\ \bibnamefont
  {Ortiz}}, \bibinfo {author} {\bibfnamefont {S.~M.}\ \bibnamefont {Teicher}},
  \bibinfo {author} {\bibfnamefont {L.}~\bibnamefont {Kautzsch}}, \bibinfo
  {author} {\bibfnamefont {P.~M.}\ \bibnamefont {Sarte}}, \bibinfo {author}
  {\bibfnamefont {N.}~\bibnamefont {Ratcliff}}, \bibinfo {author}
  {\bibfnamefont {J.}~\bibnamefont {Harter}}, \bibinfo {author} {\bibfnamefont
  {J.~P.}\ \bibnamefont {Ruff}}, \bibinfo {author} {\bibfnamefont
  {R.}~\bibnamefont {Seshadri}}, \ and\ \bibinfo {author} {\bibfnamefont
  {S.~D.}\ \bibnamefont {Wilson}},\ }\href {\doibase
  10.1103/physrevx.11.041030} {\bibfield  {journal} {\bibinfo  {journal}
  {Physical Review X}\ }\textbf {\bibinfo {volume} {11}},\ \bibinfo {pages}
  {041030} (\bibinfo {year} {2021}{\natexlab{b}})}\BibitemShut {NoStop}%
\bibitem [{\citenamefont {Schmitt}\ \emph {et~al.}(2024)\citenamefont
  {Schmitt}, \citenamefont {Biswas}, \citenamefont {Tkach}, \citenamefont
  {Fedchenko}, \citenamefont {Liu}, \citenamefont {Elmers}, \citenamefont
  {Sing}, \citenamefont {Claessen}, \citenamefont {Lee},\ and\ \citenamefont
  {Schönhense}}]{Schmitt2024}%
  \BibitemOpen
  \bibfield  {author} {\bibinfo {author} {\bibfnamefont {M.}~\bibnamefont
  {Schmitt}}, \bibinfo {author} {\bibfnamefont {D.}~\bibnamefont {Biswas}},
  \bibinfo {author} {\bibfnamefont {O.}~\bibnamefont {Tkach}}, \bibinfo
  {author} {\bibfnamefont {O.}~\bibnamefont {Fedchenko}}, \bibinfo {author}
  {\bibfnamefont {J.}~\bibnamefont {Liu}}, \bibinfo {author} {\bibfnamefont
  {H.-J.}\ \bibnamefont {Elmers}}, \bibinfo {author} {\bibfnamefont
  {M.}~\bibnamefont {Sing}}, \bibinfo {author} {\bibfnamefont {R.}~\bibnamefont
  {Claessen}}, \bibinfo {author} {\bibfnamefont {T.-L.}\ \bibnamefont {Lee}}, \
  and\ \bibinfo {author} {\bibfnamefont {G.}~\bibnamefont {Schönhense}},\
  }\href {\doibase 10.48550/ARXIV.2406.00771} {\enquote {\bibinfo {title}
  {Hybrid photoelectron momentum microscope at the soft x-ray beamline i09 of
  the diamond light source},}\ }\bibinfo {howpublished} {arXiv:2406.00771v1}
  (\bibinfo {year} {2024})\BibitemShut {NoStop}%
\bibitem [{\citenamefont {Westphal}\ \emph {et~al.}(1989)\citenamefont
  {Westphal}, \citenamefont {Bansmann}, \citenamefont {Getzlaff},\ and\
  \citenamefont {Schönhense}}]{Westphal1989}%
  \BibitemOpen
  \bibfield  {author} {\bibinfo {author} {\bibfnamefont {C.}~\bibnamefont
  {Westphal}}, \bibinfo {author} {\bibfnamefont {J.}~\bibnamefont {Bansmann}},
  \bibinfo {author} {\bibfnamefont {M.}~\bibnamefont {Getzlaff}}, \ and\
  \bibinfo {author} {\bibfnamefont {G.}~\bibnamefont {Schönhense}},\ }\href
  {\doibase 10.1103/physrevlett.63.151} {\bibfield  {journal} {\bibinfo
  {journal} {Physical Review Letters}\ }\textbf {\bibinfo {volume} {63}},\
  \bibinfo {pages} {151} (\bibinfo {year} {1989})}\BibitemShut {NoStop}%
\bibitem [{\citenamefont {Fedchenko}\ \emph {et~al.}(2024)\citenamefont
  {Fedchenko}, \citenamefont {Minár}, \citenamefont {Akashdeep}, \citenamefont
  {D’Souza}, \citenamefont {Vasilyev}, \citenamefont {Tkach}, \citenamefont
  {Odenbreit}, \citenamefont {Nguyen}, \citenamefont {Kutnyakhov},
  \citenamefont {Wind}, \citenamefont {Wenthaus}, \citenamefont {Scholz},
  \citenamefont {Rossnagel}, \citenamefont {Hoesch}, \citenamefont
  {Aeschlimann}, \citenamefont {Stadtmüller}, \citenamefont {Kläui},
  \citenamefont {Schönhense}, \citenamefont {Jungwirth}, \citenamefont
  {Hellenes}, \citenamefont {Jakob}, \citenamefont {Šmejkal}, \citenamefont
  {Sinova},\ and\ \citenamefont {Elmers}}]{Fedchenko2024}%
  \BibitemOpen
  \bibfield  {author} {\bibinfo {author} {\bibfnamefont {O.}~\bibnamefont
  {Fedchenko}}, \bibinfo {author} {\bibfnamefont {J.}~\bibnamefont {Minár}},
  \bibinfo {author} {\bibfnamefont {A.}~\bibnamefont {Akashdeep}}, \bibinfo
  {author} {\bibfnamefont {S.~W.}\ \bibnamefont {D’Souza}}, \bibinfo {author}
  {\bibfnamefont {D.}~\bibnamefont {Vasilyev}}, \bibinfo {author}
  {\bibfnamefont {O.}~\bibnamefont {Tkach}}, \bibinfo {author} {\bibfnamefont
  {L.}~\bibnamefont {Odenbreit}}, \bibinfo {author} {\bibfnamefont
  {Q.}~\bibnamefont {Nguyen}}, \bibinfo {author} {\bibfnamefont
  {D.}~\bibnamefont {Kutnyakhov}}, \bibinfo {author} {\bibfnamefont
  {N.}~\bibnamefont {Wind}}, \bibinfo {author} {\bibfnamefont {L.}~\bibnamefont
  {Wenthaus}}, \bibinfo {author} {\bibfnamefont {M.}~\bibnamefont {Scholz}},
  \bibinfo {author} {\bibfnamefont {K.}~\bibnamefont {Rossnagel}}, \bibinfo
  {author} {\bibfnamefont {M.}~\bibnamefont {Hoesch}}, \bibinfo {author}
  {\bibfnamefont {M.}~\bibnamefont {Aeschlimann}}, \bibinfo {author}
  {\bibfnamefont {B.}~\bibnamefont {Stadtmüller}}, \bibinfo {author}
  {\bibfnamefont {M.}~\bibnamefont {Kläui}}, \bibinfo {author} {\bibfnamefont
  {G.}~\bibnamefont {Schönhense}}, \bibinfo {author} {\bibfnamefont
  {T.}~\bibnamefont {Jungwirth}}, \bibinfo {author} {\bibfnamefont {A.~B.}\
  \bibnamefont {Hellenes}}, \bibinfo {author} {\bibfnamefont {G.}~\bibnamefont
  {Jakob}}, \bibinfo {author} {\bibfnamefont {L.}~\bibnamefont {Šmejkal}},
  \bibinfo {author} {\bibfnamefont {J.}~\bibnamefont {Sinova}}, \ and\ \bibinfo
  {author} {\bibfnamefont {H.-J.}\ \bibnamefont {Elmers}},\ }\href {\doibase
  10.1126/sciadv.adj4883} {\bibfield  {journal} {\bibinfo  {journal} {Science
  Advances}\ }\textbf {\bibinfo {volume} {10}},\ \bibinfo {pages} {eadj4883}
  (\bibinfo {year} {2024})}\BibitemShut {NoStop}%
\bibitem [{\citenamefont {Fedchenko}\ \emph {et~al.}(2019)\citenamefont
  {Fedchenko}, \citenamefont {Medjanik}, \citenamefont {Chernov}, \citenamefont
  {Kutnyakhov}, \citenamefont {Ellguth}, \citenamefont {Oelsner}, \citenamefont
  {Schönhense}, \citenamefont {Peixoto}, \citenamefont {Lutz}, \citenamefont
  {Min}, \citenamefont {Reinert}, \citenamefont {Däster}, \citenamefont
  {Acremann}, \citenamefont {Viefhaus}, \citenamefont {Wurth}, \citenamefont
  {Braun}, \citenamefont {Minár}, \citenamefont {Ebert}, \citenamefont
  {Elmers},\ and\ \citenamefont {Schönhense}}]{Fedchenko2019}%
  \BibitemOpen
  \bibfield  {author} {\bibinfo {author} {\bibfnamefont {O.}~\bibnamefont
  {Fedchenko}}, \bibinfo {author} {\bibfnamefont {K.}~\bibnamefont {Medjanik}},
  \bibinfo {author} {\bibfnamefont {S.}~\bibnamefont {Chernov}}, \bibinfo
  {author} {\bibfnamefont {D.}~\bibnamefont {Kutnyakhov}}, \bibinfo {author}
  {\bibfnamefont {M.}~\bibnamefont {Ellguth}}, \bibinfo {author} {\bibfnamefont
  {A.}~\bibnamefont {Oelsner}}, \bibinfo {author} {\bibfnamefont
  {B.}~\bibnamefont {Schönhense}}, \bibinfo {author} {\bibfnamefont
  {T.~R.~F.}\ \bibnamefont {Peixoto}}, \bibinfo {author} {\bibfnamefont
  {P.}~\bibnamefont {Lutz}}, \bibinfo {author} {\bibfnamefont {C.-H.}\
  \bibnamefont {Min}}, \bibinfo {author} {\bibfnamefont {F.}~\bibnamefont
  {Reinert}}, \bibinfo {author} {\bibfnamefont {S.}~\bibnamefont {Däster}},
  \bibinfo {author} {\bibfnamefont {Y.}~\bibnamefont {Acremann}}, \bibinfo
  {author} {\bibfnamefont {J.}~\bibnamefont {Viefhaus}}, \bibinfo {author}
  {\bibfnamefont {W.}~\bibnamefont {Wurth}}, \bibinfo {author} {\bibfnamefont
  {J.}~\bibnamefont {Braun}}, \bibinfo {author} {\bibfnamefont
  {J.}~\bibnamefont {Minár}}, \bibinfo {author} {\bibfnamefont
  {H.}~\bibnamefont {Ebert}}, \bibinfo {author} {\bibfnamefont {H.~J.}\
  \bibnamefont {Elmers}}, \ and\ \bibinfo {author} {\bibfnamefont
  {G.}~\bibnamefont {Schönhense}},\ }\href {\doibase 10.1088/1367-2630/aaf4cd}
  {\bibfield  {journal} {\bibinfo  {journal} {New Journal of Physics}\ }\textbf
  {\bibinfo {volume} {21}},\ \bibinfo {pages} {013017} (\bibinfo {year}
  {2019})}\BibitemShut {NoStop}%
\bibitem [{\citenamefont {Bansmann}\ \emph {et~al.}(1992)\citenamefont
  {Bansmann}, \citenamefont {Westphal}, \citenamefont {Getzlaff}, \citenamefont
  {Fegel},\ and\ \citenamefont {Schönhense}}]{Bansmann1992}%
  \BibitemOpen
  \bibfield  {author} {\bibinfo {author} {\bibfnamefont {J.}~\bibnamefont
  {Bansmann}}, \bibinfo {author} {\bibfnamefont {C.}~\bibnamefont {Westphal}},
  \bibinfo {author} {\bibfnamefont {M.}~\bibnamefont {Getzlaff}}, \bibinfo
  {author} {\bibfnamefont {F.}~\bibnamefont {Fegel}}, \ and\ \bibinfo {author}
  {\bibfnamefont {G.}~\bibnamefont {Schönhense}},\ }\href {\doibase
  10.1016/0304-8853(92)91511-q} {\bibfield  {journal} {\bibinfo  {journal}
  {Journal of Magnetism and Magnetic Materials}\ }\textbf {\bibinfo {volume}
  {104-107}},\ \bibinfo {pages} {1691} (\bibinfo {year} {1992})}\BibitemShut
  {NoStop}%
\bibitem [{\citenamefont {Nakayama}\ \emph {et~al.}(2021)\citenamefont
  {Nakayama}, \citenamefont {Li}, \citenamefont {Kato}, \citenamefont {Liu},
  \citenamefont {Wang}, \citenamefont {Takahashi}, \citenamefont {Yao},\ and\
  \citenamefont {Sato}}]{Nakayama2021}%
  \BibitemOpen
  \bibfield  {author} {\bibinfo {author} {\bibfnamefont {K.}~\bibnamefont
  {Nakayama}}, \bibinfo {author} {\bibfnamefont {Y.}~\bibnamefont {Li}},
  \bibinfo {author} {\bibfnamefont {T.}~\bibnamefont {Kato}}, \bibinfo {author}
  {\bibfnamefont {M.}~\bibnamefont {Liu}}, \bibinfo {author} {\bibfnamefont
  {Z.}~\bibnamefont {Wang}}, \bibinfo {author} {\bibfnamefont {T.}~\bibnamefont
  {Takahashi}}, \bibinfo {author} {\bibfnamefont {Y.}~\bibnamefont {Yao}}, \
  and\ \bibinfo {author} {\bibfnamefont {T.}~\bibnamefont {Sato}},\ }\href
  {\doibase 10.1103/PhysRevB.104.L161112} {\bibfield  {journal} {\bibinfo
  {journal} {Phys. Rev. B}\ }\textbf {\bibinfo {volume} {104}},\ \bibinfo
  {pages} {L161112} (\bibinfo {year} {2021})}\BibitemShut {NoStop}%
\bibitem [{\citenamefont {Azoury}\ \emph {et~al.}(2023)\citenamefont {Azoury},
  \citenamefont {von Hoegen}, \citenamefont {Su}, \citenamefont {Oh},
  \citenamefont {Holder}, \citenamefont {Tan}, \citenamefont {Ortiz},
  \citenamefont {Capa~Salinas}, \citenamefont {Wilson}, \citenamefont {Yan},\
  and\ \citenamefont {Gedik}}]{Azoury2023}%
  \BibitemOpen
  \bibfield  {author} {\bibinfo {author} {\bibfnamefont {D.}~\bibnamefont
  {Azoury}}, \bibinfo {author} {\bibfnamefont {A.}~\bibnamefont {von Hoegen}},
  \bibinfo {author} {\bibfnamefont {Y.}~\bibnamefont {Su}}, \bibinfo {author}
  {\bibfnamefont {K.~H.}\ \bibnamefont {Oh}}, \bibinfo {author} {\bibfnamefont
  {T.}~\bibnamefont {Holder}}, \bibinfo {author} {\bibfnamefont
  {H.}~\bibnamefont {Tan}}, \bibinfo {author} {\bibfnamefont {B.~R.}\
  \bibnamefont {Ortiz}}, \bibinfo {author} {\bibfnamefont {A.}~\bibnamefont
  {Capa~Salinas}}, \bibinfo {author} {\bibfnamefont {S.~D.}\ \bibnamefont
  {Wilson}}, \bibinfo {author} {\bibfnamefont {B.}~\bibnamefont {Yan}}, \ and\
  \bibinfo {author} {\bibfnamefont {N.}~\bibnamefont {Gedik}},\ }\href
  {\doibase 10.1073/pnas.2308588120} {\bibfield  {journal} {\bibinfo  {journal}
  {Proceedings of the National Academy of Sciences}\ }\textbf {\bibinfo
  {volume} {120}},\ \bibinfo {pages} {e2308588120} (\bibinfo {year}
  {2023})}\BibitemShut {NoStop}%
\bibitem [{\citenamefont {Tkach}\ \emph {et~al.}(2024)\citenamefont {Tkach},
  \citenamefont {Fragkos}, \citenamefont {Nguyen}, \citenamefont {Chernov},
  \citenamefont {Scholz}, \citenamefont {Wind}, \citenamefont {Babenkov},
  \citenamefont {Fedchenko}, \citenamefont {Lytvynenko}, \citenamefont
  {Zimmer}, \citenamefont {Hloskovskii}, \citenamefont {Kutnyakhov},
  \citenamefont {Pressacco}, \citenamefont {Dilling}, \citenamefont
  {Bruckmeier}, \citenamefont {Heber}, \citenamefont {Scholz}, \citenamefont
  {Sobota}, \citenamefont {Koralek}, \citenamefont {Sirica}, \citenamefont
  {Kallmayer}, \citenamefont {Hoesch}, \citenamefont {Schlueter}, \citenamefont
  {Odnodvorets}, \citenamefont {Mairesse}, \citenamefont {Rossnagel},
  \citenamefont {Elmers}, \citenamefont {Beaulieu},\ and\ \citenamefont
  {Schoenhense}}]{Tkach2024}%
  \BibitemOpen
  \bibfield  {author} {\bibinfo {author} {\bibfnamefont {O.}~\bibnamefont
  {Tkach}}, \bibinfo {author} {\bibfnamefont {S.}~\bibnamefont {Fragkos}},
  \bibinfo {author} {\bibfnamefont {Q.}~\bibnamefont {Nguyen}}, \bibinfo
  {author} {\bibfnamefont {S.}~\bibnamefont {Chernov}}, \bibinfo {author}
  {\bibfnamefont {M.}~\bibnamefont {Scholz}}, \bibinfo {author} {\bibfnamefont
  {N.}~\bibnamefont {Wind}}, \bibinfo {author} {\bibfnamefont {S.}~\bibnamefont
  {Babenkov}}, \bibinfo {author} {\bibfnamefont {O.}~\bibnamefont {Fedchenko}},
  \bibinfo {author} {\bibfnamefont {Y.}~\bibnamefont {Lytvynenko}}, \bibinfo
  {author} {\bibfnamefont {D.}~\bibnamefont {Zimmer}}, \bibinfo {author}
  {\bibfnamefont {A.}~\bibnamefont {Hloskovskii}}, \bibinfo {author}
  {\bibfnamefont {D.}~\bibnamefont {Kutnyakhov}}, \bibinfo {author}
  {\bibfnamefont {F.}~\bibnamefont {Pressacco}}, \bibinfo {author}
  {\bibfnamefont {J.}~\bibnamefont {Dilling}}, \bibinfo {author} {\bibfnamefont
  {L.}~\bibnamefont {Bruckmeier}}, \bibinfo {author} {\bibfnamefont
  {M.}~\bibnamefont {Heber}}, \bibinfo {author} {\bibfnamefont
  {F.}~\bibnamefont {Scholz}}, \bibinfo {author} {\bibfnamefont
  {J.}~\bibnamefont {Sobota}}, \bibinfo {author} {\bibfnamefont
  {J.}~\bibnamefont {Koralek}}, \bibinfo {author} {\bibfnamefont
  {N.}~\bibnamefont {Sirica}}, \bibinfo {author} {\bibfnamefont
  {M.}~\bibnamefont {Kallmayer}}, \bibinfo {author} {\bibfnamefont
  {M.}~\bibnamefont {Hoesch}}, \bibinfo {author} {\bibfnamefont
  {C.}~\bibnamefont {Schlueter}}, \bibinfo {author} {\bibfnamefont {L.~V.}\
  \bibnamefont {Odnodvorets}}, \bibinfo {author} {\bibfnamefont
  {Y.}~\bibnamefont {Mairesse}}, \bibinfo {author} {\bibfnamefont
  {K.}~\bibnamefont {Rossnagel}}, \bibinfo {author} {\bibfnamefont {H.~J.}\
  \bibnamefont {Elmers}}, \bibinfo {author} {\bibfnamefont {S.}~\bibnamefont
  {Beaulieu}}, \ and\ \bibinfo {author} {\bibfnamefont {G.}~\bibnamefont
  {Schoenhense}},\ }\href {\doibase 10.48550/ARXIV.2401.10084} {\enquote
  {\bibinfo {title} {Multi-mode front lens for momentum microscopy: Part ii
  experiments},}\ }\bibinfo {howpublished} {arXiv:2401.10084v2} (\bibinfo
  {year} {2024})\BibitemShut {NoStop}%
\bibitem [{\citenamefont {Li}\ \emph {et~al.}(2024)\citenamefont {Li},
  \citenamefont {Kim},\ and\ \citenamefont {Kee}}]{Li2024}%
  \BibitemOpen
  \bibfield  {author} {\bibinfo {author} {\bibfnamefont {H.}~\bibnamefont
  {Li}}, \bibinfo {author} {\bibfnamefont {Y.~B.}\ \bibnamefont {Kim}}, \ and\
  \bibinfo {author} {\bibfnamefont {H.-Y.}\ \bibnamefont {Kee}},\ }\href
  {\doibase 10.1103/physrevlett.132.146501} {\bibfield  {journal} {\bibinfo
  {journal} {Physical Review Letters}\ }\textbf {\bibinfo {volume} {132}},\
  \bibinfo {pages} {146501} (\bibinfo {year} {2024})}\BibitemShut {NoStop}%
\bibitem [{\citenamefont {Xiao}\ \emph {et~al.}(2023)\citenamefont {Xiao},
  \citenamefont {Lin}, \citenamefont {Li}, \citenamefont {Zheng}, \citenamefont
  {Francoual}, \citenamefont {Plueckthun}, \citenamefont {Xia}, \citenamefont
  {Qiu}, \citenamefont {Zhang}, \citenamefont {Guo}, \citenamefont {Feng},\
  and\ \citenamefont {Peng}}]{Xiao2023}%
  \BibitemOpen
  \bibfield  {author} {\bibinfo {author} {\bibfnamefont {Q.}~\bibnamefont
  {Xiao}}, \bibinfo {author} {\bibfnamefont {Y.}~\bibnamefont {Lin}}, \bibinfo
  {author} {\bibfnamefont {Q.}~\bibnamefont {Li}}, \bibinfo {author}
  {\bibfnamefont {X.}~\bibnamefont {Zheng}}, \bibinfo {author} {\bibfnamefont
  {S.}~\bibnamefont {Francoual}}, \bibinfo {author} {\bibfnamefont
  {C.}~\bibnamefont {Plueckthun}}, \bibinfo {author} {\bibfnamefont
  {W.}~\bibnamefont {Xia}}, \bibinfo {author} {\bibfnamefont {Q.}~\bibnamefont
  {Qiu}}, \bibinfo {author} {\bibfnamefont {S.}~\bibnamefont {Zhang}}, \bibinfo
  {author} {\bibfnamefont {Y.}~\bibnamefont {Guo}}, \bibinfo {author}
  {\bibfnamefont {J.}~\bibnamefont {Feng}}, \ and\ \bibinfo {author}
  {\bibfnamefont {Y.}~\bibnamefont {Peng}},\ }\href {\doibase
  10.1103/physrevresearch.5.l012032} {\bibfield  {journal} {\bibinfo  {journal}
  {Physical Review Research}\ }\textbf {\bibinfo {volume} {5}},\ \bibinfo
  {pages} {L012032} (\bibinfo {year} {2023})}\BibitemShut {NoStop}%
\bibitem [{\citenamefont {Feng}\ \emph {et~al.}(2023)\citenamefont {Feng},
  \citenamefont {Zhao}, \citenamefont {Luo}, \citenamefont {Yang},
  \citenamefont {Fang}, \citenamefont {Yang}, \citenamefont {Gao},
  \citenamefont {Zhou},\ and\ \citenamefont {Zheng}}]{Feng2023}%
  \BibitemOpen
  \bibfield  {author} {\bibinfo {author} {\bibfnamefont {X.~Y.}\ \bibnamefont
  {Feng}}, \bibinfo {author} {\bibfnamefont {Z.}~\bibnamefont {Zhao}}, \bibinfo
  {author} {\bibfnamefont {J.}~\bibnamefont {Luo}}, \bibinfo {author}
  {\bibfnamefont {J.}~\bibnamefont {Yang}}, \bibinfo {author} {\bibfnamefont
  {A.~F.}\ \bibnamefont {Fang}}, \bibinfo {author} {\bibfnamefont {H.~T.}\
  \bibnamefont {Yang}}, \bibinfo {author} {\bibfnamefont {H.~J.}\ \bibnamefont
  {Gao}}, \bibinfo {author} {\bibfnamefont {R.}~\bibnamefont {Zhou}}, \ and\
  \bibinfo {author} {\bibfnamefont {G.-q.}\ \bibnamefont {Zheng}},\ }\href
  {\doibase 10.1038/s41535-023-00555-w} {\bibfield  {journal} {\bibinfo
  {journal} {npj Quantum Materials}\ }\textbf {\bibinfo {volume} {8}},\
  \bibinfo {pages} {23} (\bibinfo {year} {2023})}\BibitemShut {NoStop}%
\bibitem [{\citenamefont {Frassineti}\ \emph {et~al.}(2023)\citenamefont
  {Frassineti}, \citenamefont {Bonfà}, \citenamefont {Allodi}, \citenamefont
  {Garcia}, \citenamefont {Cong}, \citenamefont {Ortiz}, \citenamefont
  {Wilson}, \citenamefont {De~Renzi}, \citenamefont {Mitrović},\ and\
  \citenamefont {Sanna}}]{Frassineti2023}%
  \BibitemOpen
  \bibfield  {author} {\bibinfo {author} {\bibfnamefont {J.}~\bibnamefont
  {Frassineti}}, \bibinfo {author} {\bibfnamefont {P.}~\bibnamefont {Bonfà}},
  \bibinfo {author} {\bibfnamefont {G.}~\bibnamefont {Allodi}}, \bibinfo
  {author} {\bibfnamefont {E.}~\bibnamefont {Garcia}}, \bibinfo {author}
  {\bibfnamefont {R.}~\bibnamefont {Cong}}, \bibinfo {author} {\bibfnamefont
  {B.~R.}\ \bibnamefont {Ortiz}}, \bibinfo {author} {\bibfnamefont {S.~D.}\
  \bibnamefont {Wilson}}, \bibinfo {author} {\bibfnamefont {R.}~\bibnamefont
  {De~Renzi}}, \bibinfo {author} {\bibfnamefont {V.~F.}\ \bibnamefont
  {Mitrović}}, \ and\ \bibinfo {author} {\bibfnamefont {S.}~\bibnamefont
  {Sanna}},\ }\href {\doibase 10.1103/physrevresearch.5.l012017} {\bibfield
  {journal} {\bibinfo  {journal} {Physical Review Research}\ }\textbf {\bibinfo
  {volume} {5}},\ \bibinfo {pages} {L012017} (\bibinfo {year}
  {2023})}\BibitemShut {NoStop}%
\bibitem [{\citenamefont {Brinkman}\ \emph {et~al.}(2024)\citenamefont
  {Brinkman}, \citenamefont {Tan}, \citenamefont {Brekke}, \citenamefont
  {Mathisen}, \citenamefont {Finnseth}, \citenamefont {Schenk}, \citenamefont
  {Hagiwara}, \citenamefont {Huang}, \citenamefont {Buck}, \citenamefont
  {Kallane}, \citenamefont {Hoesch}, \citenamefont {Rossnagel}, \citenamefont
  {Ou~Yang}, \citenamefont {Lin}, \citenamefont {Shu}, \citenamefont {Chen},
  \citenamefont {Tusche},\ and\ \citenamefont {Bentmann}}]{Brinkman2024}%
  \BibitemOpen
  \bibfield  {author} {\bibinfo {author} {\bibfnamefont {S.~S.}\ \bibnamefont
  {Brinkman}}, \bibinfo {author} {\bibfnamefont {X.~L.}\ \bibnamefont {Tan}},
  \bibinfo {author} {\bibfnamefont {B.}~\bibnamefont {Brekke}}, \bibinfo
  {author} {\bibfnamefont {A.~C.}\ \bibnamefont {Mathisen}}, \bibinfo {author}
  {\bibfnamefont {O.}~\bibnamefont {Finnseth}}, \bibinfo {author}
  {\bibfnamefont {R.~J.}\ \bibnamefont {Schenk}}, \bibinfo {author}
  {\bibfnamefont {K.}~\bibnamefont {Hagiwara}}, \bibinfo {author}
  {\bibfnamefont {M.-J.}\ \bibnamefont {Huang}}, \bibinfo {author}
  {\bibfnamefont {J.}~\bibnamefont {Buck}}, \bibinfo {author} {\bibfnamefont
  {M.}~\bibnamefont {Kallane}}, \bibinfo {author} {\bibfnamefont
  {M.}~\bibnamefont {Hoesch}}, \bibinfo {author} {\bibfnamefont
  {K.}~\bibnamefont {Rossnagel}}, \bibinfo {author} {\bibfnamefont {K.-H.}\
  \bibnamefont {Ou~Yang}}, \bibinfo {author} {\bibfnamefont {M.-T.}\
  \bibnamefont {Lin}}, \bibinfo {author} {\bibfnamefont {G.-J.}\ \bibnamefont
  {Shu}}, \bibinfo {author} {\bibfnamefont {Y.-J.}\ \bibnamefont {Chen}},
  \bibinfo {author} {\bibfnamefont {C.}~\bibnamefont {Tusche}}, \ and\ \bibinfo
  {author} {\bibfnamefont {H.}~\bibnamefont {Bentmann}},\ }\href {\doibase
  10.1103/physrevlett.132.196402} {\bibfield  {journal} {\bibinfo  {journal}
  {Physical Review Letters}\ }\textbf {\bibinfo {volume} {132}},\ \bibinfo
  {pages} {196402} (\bibinfo {year} {2024})}\BibitemShut {NoStop}%
\end{thebibliography}
%merlin.mbs apsrev4-1.bst 2010-07-25 4.21a (PWD, AO, DPC) hacked
%Control: key (0)
%Control: author (72) initials jnrlst
%Control: editor formatted (1) identically to author
%Control: production of article title (-1) disabled
%Control: page (0) single
%Control: year (1) truncated
%Control: production of eprint (0) enabled
%

\end{document}